\documentclass{jfm}

\usepackage{graphicx}
\usepackage{epsfig}
\usepackage{newtxtext}
\usepackage{newtxmath}
\usepackage{natbib}
\usepackage{hyperref}
\usepackage{upgreek}
\usepackage{bm}
\hypersetup{
    colorlinks = true,
    urlcolor   = blue,
    citecolor  = black,
}

\newcommand{\RomanNumeralCaps}[1]
\linenumbers

\title{Elastic wakes mediate collective viscoelastic fluid-structure interactions in side-by-side cantilever arrays}

\author{Arisa Yokokoji,
Amy Q.\ Shen,
 \and Simon J.\ Haward \corresp{\email{simon.haward@oist.jp}}}

\affiliation{Micro/Bio/Nanofluidics Unit, Okinawa Institute of Science and Technology Graduate University, Onna, Okinawa 904-0495, Japan.}

\begin{document}
\maketitle

\begin{abstract}
Fluid–structure interaction (FSI) in viscoelastic flows past deformable structures at low Reynolds numbers remains poorly understood, despite its relevance to biological systems such as cilia and flagella, and to engineered microsystems. We investigate viscoelastic FSI in side-by-side flexible cantilever arrays using a bottom-up approach that systematically varies the number of cantilevers and the rheology of the test fluid, comparing weakly shear-thinning (WS) and highly shear-thinning (HS) polyethylene oxide solutions. For both fluids, with increasing Weissenberg number (Wi), an elongated elastic wake develops behind a single isolated cantilever. For multiple cantilevers, the WS fluid undergoes a transition at a critical Weissenberg number ($\text{Wi}^{*}$) from separated to merged elastic wakes, accompanied by the emergence of a divergent flow field and coordinated inward spanwise cantilever deflection. The critical $\text{Wi}^{*}$ increases as the number of cantilevers in the array is increased from two to three, demonstrating a strong dependence of the onset on the array configuration. For the HS fluid, wake merger, flow divergence, and inward cantilever deflection are suppressed across the full range of Wi investigated, despite comparable elasticity and the formation of elastic wakes. This contrast shows that elasticity alone promotes wake formation but is insufficient to produce the collective instability, which instead requires both shear-thinning and interactions between neighboring cantilevers. These findings demonstrate that viscoelastic FSI in flexible arrays is governed by the combined effects of fluid elasticity, shear-thinning, and geometric configuration, with elastic wake interactions mediating the collective instability and linking local elastic wakes to array-scale structural response.
\end{abstract}

%{\bf MSC Codes }  {\it(Optional)} Please enter your MSC Codes here

\section{Introduction}
\label{sec:intro}

Fluid-structure interaction (FSI) between Newtonian fluids and deformable or displaceable structures has been extensively studied over the past few decades \citep{Shelley2011}. At high Reynolds numbers ($\text{Re} = Ul/\nu \gg 1$, where $U$ and $l$ are the characteristic flow velocity and length scale, respectively, and $\nu$ is the kinematic viscosity), instabilities arising from fluid inertia give rise to rich structural responses, observable in natural phenomena such as the reconfiguration of trees and plants in strong wind, and of sea anemones in the ocean \citep{Vogel1996}. A canonical example is vortex-induced vibration around a bluff body, such as a cylindrical cantilever \citep{Bearman1984, Sarpkaya2004, Williamson2008,Huera2025}.

In contrast to Newtonian fluids, viscoelastic fluids (e.g., polymeric or micellar solutions) can develop flow instabilities even at low Reynolds numbers ($\text{Re} \ll 1$) due to nonlinearities arising from fluid elasticity, as quantified by the Weissenberg number $\text{Wi} = \tau U/ l$, where $\tau$ is the characteristic time of the fluid. Such ``purely-elastic'' instabilities arise in the regime of high elasticity number $\text{El}=\text{Wi}/\text{Re} > 1$ \citep{Datta2022}. However, FSI in viscoelastic fluids at low Re and high Wi (or high El) remains largely unexplored, despite the considerable relevance of such regimes to biological and mechanical systems at small length scales. Cilia and flagella, for example, often operate in viscoelastic mucus where non-Newtonian rheological properties (i.e., elasticity, and also shear-thinning) fundamentally influence their dynamics and ability to function \citep{Chen1978,Li2017,Cicuta2020,Haward_FSI, Schneiter2021}. A deeper understanding of FSI in viscoelastic fluids also has direct implications for engineered microsystems, including micromixing and micro-object manipulation \citep{Toonder2013,Musharaf2024}.

For laboratory study of viscoelastic flows there are two basic routes to increase Wi relative to Re and thus enhance El: (1) employ a highly viscous fluid with a large value for $\tau$, or (2) employ flow geometries with small characteristic lengthscales $l$, \textit{viz}., use microfluidic devices. In the past, both approaches have been used to examine high-El flows past rigidly-fixed circular cylinders, and to a lesser extent around cantilevered, or flexibly-mounted, circular cylinders. 

While for Newtonian flow at low Re, the flow field around a symmetric body, such as a rigidly-fixed cylinder, remains both fore-aft and laterally symmetric, viscoelasticity of the fluid can break this symmetry in a variety of ways. For $\text{Wi} \gtrsim 1$, polymers can become elongated and aligned by the strong extensional kinematics around the leading and trailing stagnation points of a cylinder, leading to the formation of a highly extended downstream wake of low flow velocity and high elastic tensile stress that breaks the fore-aft symmetry \citep{Chilcott1988,Francois2008,Nolan2016,Haward2018}. Furthermore, the generation of elastic tensile stress along the curving streamlines around the cylinder induces a hoop stress that can result in purely elastic instabilities  \citep{Mckinley1996,Pakdel1996}. These can take the form of streamline distortions near the leading and trailing edges of the cylinder and/or the onset of three-dimensional flow \citep{Mckinley1993,Shiang2000,Moss2010, Ribeiro2014, Haward2018, Haward2021}. It has further been shown how shear-thinning properties of viscoelastic fluids influence the onset and development of instability. Shear-thinning can promote the development of a steady laterally asymmetric flow state whereby fluid flows preferentially around one side of the cylinder or the other \citep{Haward2019,Haward2020,Varchanis2020,Khan2021,Haward2021,Spyridakis2024}. Since the onset of laterally asymmetric flow represents a severe modification of the flow field, it naturally also modifies the stress distribution around, and downstream of, the cylinder \citep{Haward2019,Varchanis2020,Khan2021}. These studies demonstrate that viscoelasticity can fundamentally alter the flow topology around individual obstacles. However, how elastic wakes interact between neighboring flexible structures, and how such interactions influence their coupled dynamics, remains poorly understood.

Within this framework, a few studies have examined FSI between viscoelastic fluid and the dynamics of single or multiple deformable cylinders. One notable investigation conducted by \citet{Dey2018} examined the flow of a viscoelastic and highly shear-thinning wormlike micellar (WLM) solution within a straight channel containing a flexible polydimethylsiloxane (PDMS) cylinder pinned at both ends. The channel and flexible cylinder were of millimetric dimensions, but the high viscosity and long characteristic time $\tau$ of the WLM solution ensured that $\text{Re} \ll 1$ and that the flow was dominated by elasticity. The researchers tracked the deformation of the flexible cylinder, performed particle imaging velocimetry (PIV) tests, and flow-induced birefringence (FIB) tests. As the imposed flow velocity was incremented such that a critical Weissenberg number $\text{Wi}_1 \approx 100$ was exceeded, the flexible cylinder commenced one-dimensional (1D) oscillations in the flow direction, following the periodic growth and rupture of the elastic wake that formed behind it. Similar behavior was also observed in a flexible PDMS sheet by \citet{Dey2017}. Beyond a second critical Weissenberg number ($\text{Wi}_2 \approx 220$), two-dimensional (2D)  oscillations of the cylinder were observed. The 2D time-dependent motion was associated with a significantly asymmetric wake, vortex structures immediately behind the cylinder from PIV tests, and the time-dependent evolution of the elastic wake length from FIB tests. These findings underscore connections between the flow field, the elastic wake induced by the stretching of WLMs in the vicinity of the cylinder, and the time-dependent displacement of the flexible structure.

Similar behavior was reported by \citet{Hopkins2020} also using a viscoelastic WLM solution. They employed cantilevered glass cylinders of diameter $16~\upmu$m and height 1.94~mm contained within glass microchannels of width $400~\upmu$m and height 2~mm. Over a wide range of imposed Wi, they investigated the responses of both a single isolated cantilever and also of two cantilevers separated by a distance of $400~\upmu$m along the flow direction. In their investigation of the single cantilever, the authors observed a modest 2D displacement of the cantilever and the development of an asymmetric elastic wake above the first critical Weissenberg number $\text{Wi}_1 \approx 75$. Above the second critical Weissenberg number $\text{Wi}_2 \approx 145$, the displacement of the cantilever exhibited more pronounced and time-dependent 2D motion. Concurrently, the elastic wake trailing the cantilever became increasingly asymmetric and time-dependent. Above $\text{Wi}_2$, they conducted an analysis of the power spectral density (PSD) derived from the time series of the cantilever's position in the flow direction. This analysis unveiled a distinct peak at a frequency corresponding to the inverse of the Maxwell relaxation time $\tau_{Maxwell}$, determined from small-angle oscillatory shear rheometry. The dominant frequency indicates a correlation between the periodic movement of the cantilever and the lifetime of the wormlike micelles, as proposed by \citet{Dey2018}. In the high $\mathrm{Wi}$ regime ($\text{Wi}\gg \text{Wi}_2$), a steep power-law decay emerged from the PSD of the time-dependent streamwise motion of the cantilever, with the slope of $\approx -3.5$ being indicative of elastic turbulence \citep{Groisman2000,Steinberg2021,Datta2022}. This analysis emphasized that the cantilever's motion was primarily driven by the elastic stress exerted by the WLMs within the solution. In the case of the twin-cantilever configuration, for $\text{Wi}\geq \text{Wi}_2$ the periodic oscillatory motion of the two cantilevers was found to be highly correlated, particularly in the spanwise direction transverse to the flow~\citep{Hopkins2020}. FIB results revealed a highly localized birefringent elastic wake originating from the upstream cantilever and connecting with the downstream cantilever. The study posited that the time-dependent variations in elastic stress generated at the upstream cantilever were transmitted through its wake as an elastic wave, impacting the motion of the downstream cantilever and facilitating their synchronization.

Building upon the findings from simpler systems, \citet{Blois2023} conducted a study involving a more complex multiple cantilever array system within a straight channel. They created a micro-cantilever array structure (effectively a ``micro-canopy'', see e.g.,~\citet{Nepf2012}) formed from 14,400 individual cantilevered posts, each with a diameter of just $2~\upmu$m, a height of $48~\upmu$m, and spaced $12~\upmu$m apart on a square grid. They examined the flow behavior and post motion using a glycerol-water solution as a Newtonian reference fluid and sodium hyaluronic acid solutions as non-Newtonian fluids. Remarkably, the study revealed a distinctive wave pattern forming at the surface of the canopy (i.e., at the free ends of the cantilevers), but only in the case of the non-Newtonian fluid beyond a critical Weissenberg number $\text{Wi}^* \approx 100$. This wave pattern, characterized by a constant angle $\beta \approx 30^{\circ}$ relative to the flow direction, propagated downstream. The authors attribute this emergent Monami-like wave pattern to the interaction between the viscoelastic fluid and the cantilever array, particularly the effect of an elastic analogue of the Kelvin-Helmholtz instability caused by a strong shear layer at the canopy surface.

The studies described above reveal two significant knowledge gaps in viscoelastic FSI. First, the relative roles of fluid elasticity and shear-thinning in determining the flow field and structural response remain difficult to disentangle, since both rheological effects are typically present simultaneously. Second, the connection between elastic wakes generated by isolated structures, the motion of individual cantilevers, and the collective dynamics observed in flexible arrays remains unclear.
 To address these gaps, we adopt a bottom-up approach by systematically studying viscoelastic FSI in line-arrays of cantilevers arranged spanwise to the flow. By varying the number of cantilevers and comparing fluids with contrasting shear-thinning behaviour, we examine how elasticity, shear-thinning, and geometric configuration together govern the coupled flow and structural response. Our results reveal a collective viscoelastic fluid--structure instability in which neighbouring elastic wakes merge, producing a divergent-flow state and inward spanwise cantilever deflection. The onset of this instability depends on the number of cantilevers in the array and is suppressed in the highly shear-thinning fluid, indicating that collective wake interactions control the instability threshold while rheology and geometry determine whether the merged-wake state can develop.

\begin{figure*}
\centerline{\includegraphics[scale=0.28]{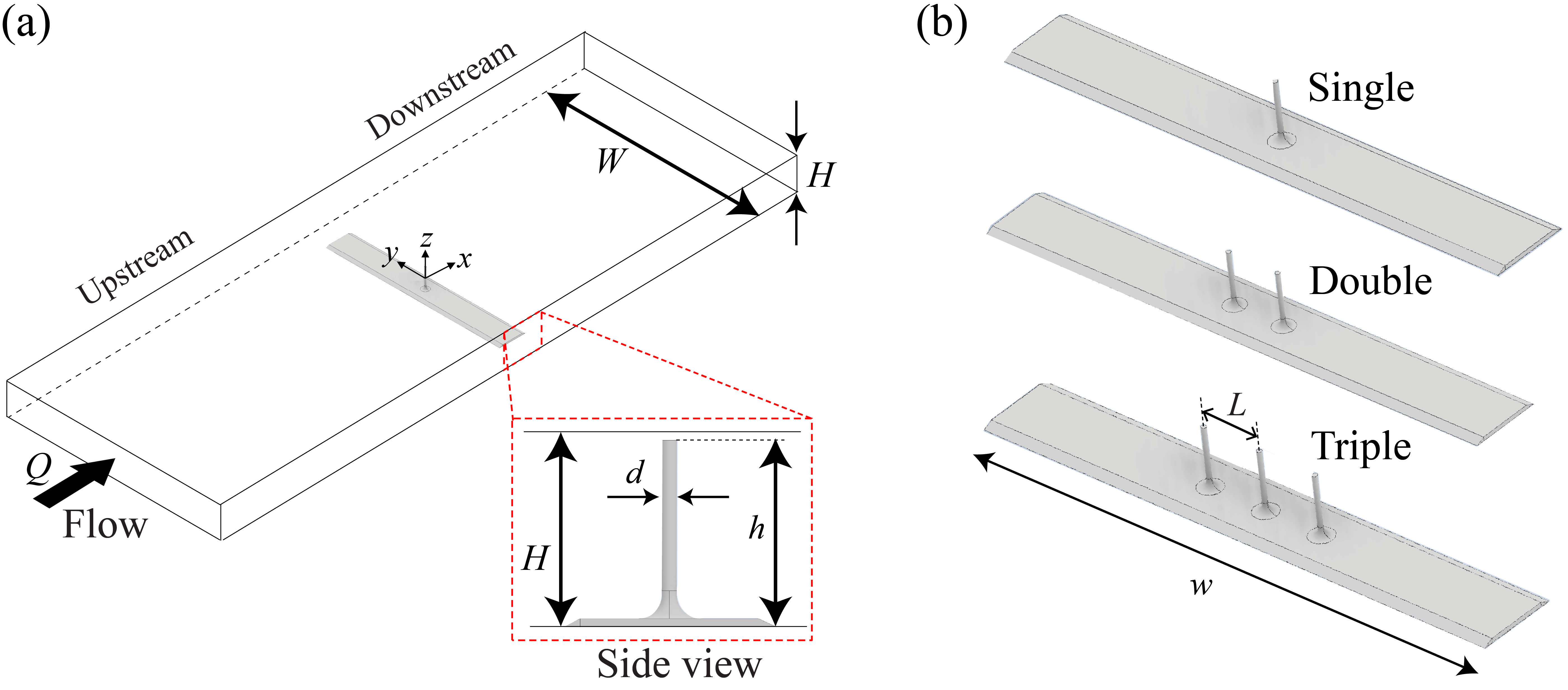}}
  %\captionsetup{style=capcenter}
  \caption{Experimental microfluidic platform used for the viscoelastic fluid--structure interaction experiments.  (a) Schematic illustration of the flow geometry. A straight, rectangular cross-section channel of width $W=1$~mm and height $H=100~\upmu$m is fabricated in glass by selective laser-induced etching. Circular cantilevers of diameter $d=8~\upmu$m and height $h=96~\upmu$m (shaded gray) are 3D printed inside the glass channel by two-photon polymerization of a polymeric resin. The origin of coordinates is taken at the center of the glass channel, with $x$ being the flow direction, and the line arrays of cantilevers are located along $x=0$. $Q$ indicates the volumetric flow rate regulated by a syringe pump. (b) Three cantilever array configurations investigated in the present study: single, double, and triple cantilevers. Adjacent cantilevers are separated by a center-to-center distance of $L=96~\upmu$m and are supported by a common base of total width $w$ (defined in the main text).}
  \label{fgr_channel}
\end{figure*}

\section{Experimental Methods}

\subsection{Microchannel and cantilever array fabrication}

A straight channel of rectangular cross-section of width $W=1$~mm and height $H=100~\upmu$m and length 15~mm, open on one side, is designed using the 3D design software Rhinoceros V5, and is fabricated in fused silica glass by using Selective Laser-induced Etching (SLE) on a Lightfab SLE printer (LightFab GmbH, Germany) \citep{Gottmann2012,Meineke2016,Burshtein2019}.
The polymeric cantilever arrays are fabricated from commercial IP-Dip photoresist (acrylate-based polymer, Nanoscribe). The resin is deposited inside the open glass microchannel and polymerized by using Photonic Professional GT2 (Nanoscribe, Germany), which enables direct laser writing via two-photon polymerization \citep{Kawata2001,Buckmann2012}. As shown schematically in Fig.~\ref{fgr_channel}(a), the cantilever arrays are positioned halfway along the length of the channel (at $x=0$) and aligned across the channel (along $y$). The various cantilever arrays (single, double and triple) are shown in Fig.~\ref{fgr_channel}(b). They are designed on the top of a base with a thickness of $4~\upmu$m through the $z$-direction, a length of $96~\upmu$m in the $x$-direction, a width through $y$ of $864~\upmu$m for the single and triple cantilever arrays and $960~\upmu$m for the double cantilever array. The $x$-$y$ dimensions of the base help to ensure that the cantilevers are printed in the center of the straight microchannel, and the thickness of the base mitigates the effects of channel floor roughness on the fabrication process. Each cantilever has a diameter $d=8~\upmu$m, height $h= 96~\upmu$m. In the case of the multiple cantilevers, the cantilevers are separated by a center-to-center distance of $L=96~\upmu$m in the $y$-direction.

After printing the arrays inside the open microchannel, the excess unpolymerized resin is carefully rinsed away following the protocol described by \citet{Blois2023}. Finally, the open glass channel containing the array is sealed against a glass slide and bonded using two-part epoxy resin to make a sealed and enclosed flow geometry. Both ends of the channel are fitted with connectors for fluid inlet and outlet.

 When polymerised, the IP-Dip photoresist resin has an estimated Young's modulus in the range $0.27 < E < 1.8~\text{GPa}$, depending on printing conditions (i.e., applied laser power and scanning speed) \citep{Schweiger2022}. The resulting bending stiffness, would be in the range $0.06 < EI < 0.37~\text{pN~m}^2$, where $I = \pi d^4/64$ is the second moment of area of a circle. Based on the printing parameters used here, the bending stiffness is expected to be at the lower end of this estimated range.

\subsection{Sample preparation and rheological characterization}

\begin{figure*}
%\centering
  \centerline{\includegraphics[scale=0.265]{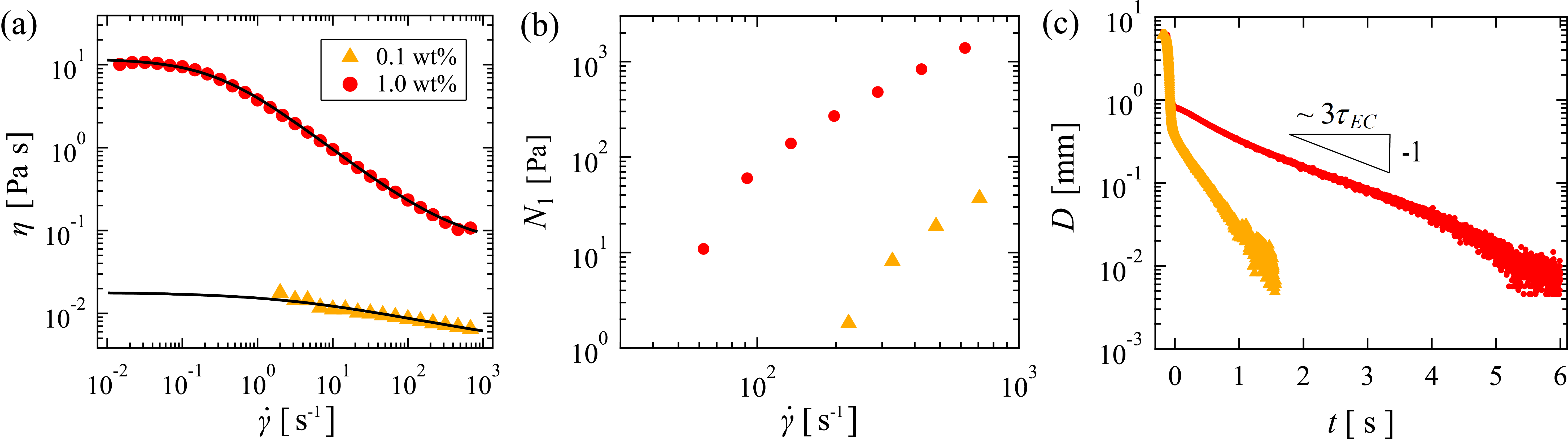}}
  %\captionsetup{style=capcenter}
  \caption{Rheological characterization of the weakly shear-thinning (WS, $c=0.1$ wt\%) and highly shear-thinning (HS, $c=1.0$ wt\%) poly(ethylene oxide) (PEO) solutions. (a) Shear viscosity $\eta$ as a function of shear rate $\dot{\gamma}$ and (b) the first normal stress difference $N_1$ as a function of $\dot{\gamma}$ obtained using a stress-controlled rotational rheometer (MCR 502, Anton-Paar). The solid black lines in (a) represent the fit of the Carreau-Yasuda model (Eq.~\ref{eq:carreau}). (c) Temporal evolution of the filament diameter $D(t)$ during capillary-driven thinning using a capillary breakup extensional rheometer (CaBER, Thermo Haake). The exponential thinning regime is used to determine the elastocapillary thinning time $\tau_{EC}$ reported in Table~\ref{tab:carreau}. All measurements are performed at $25^{\circ}$C. }
  \label{fgr:Rheology}
\end{figure*}

Polymeric solutions at concentrations of $c= 0.1$ and $c = 1.0$~wt\% are prepared by dissolving poly(ethylene oxide) (PEO, $M_v = 8\times10^6~\text{g~mol}^{-1}$, Sigma Aldrich) in an aqueous mixture of sodium iodide (NaI, 26~wt\%, Wako) and glycerol (14~wt\%, Sigma Aldrich). Each solution is mixed on a roller for at least four days to ensure homogeneity. The NaI/glycerol aqueous solvent system (with viscosity $\eta_s = 1.78~\text{mPa s}$, density $\rho = 1290~\text{kg~m}^{-3}$) is formulated to achieve a refractive index RI = 1.40 at a wavelength of 589~nm, sufficiently close to that of the polymeric cantilever array (RI = 1.54 at 589~nm) to mitigate the effects of optical distortion on flow velocimetry measurements.

As well as the polymeric test solutions, we employ two Newtonian reference fluids. One of the Newtonian fluids is a refractive index matched aqueous solution of 45~wt\% NaI in order to facilitate flow velocimetry. The other is an aqueous solution of 70~wt\% glycerol designed to match the zero shear rate viscosity of the $c= 0.1$~wt\% PEO solution, see below. 
\subsubsection{\label{sec:shearrheo}Shear Rheology}

Steady shear flow measurements on the polymeric fluids are carried out at $T$ = 25 $^{\circ}$C using an Anton Paar MCR 502 stress-controlled rheometer fitted with a 50~mm diameter $1^{\circ}$ angle cone-plate geometry. The steady shear viscosity of the test solutions over the shear rate range of $0.01 < \dot{\gamma} < 1000~\text{s}^{-1}$ is shown in Fig.~\ref{fgr:Rheology}(a). With increasing shear rates, the 0.1~wt\% PEO solution is seen to be mildly shear-thinning, with an almost constant viscosity $\eta \approx 0.01~\text{Pa s}$ over the range of tested shear rates. In contrast, the 1~wt\% PEO solution has a much higher viscosity $\eta \approx 10~\text{Pa s}$ at low shear rates and exhibits around two orders of magnitude of shear-thinning as the rate increases. Both flow curves shown in Fig.~\ref{fgr:Rheology}(a) are well described by the Carreau-Yasuda (CY) model (solid lines), described below:

\begin{equation}
\label{eq:carreau}
\eta = \eta_{\infty} + \frac{\eta_{0}-\eta_{\infty}}{[1+(\dot{\gamma}/\dot\gamma_{c})^a]^{(1-n)/a}},
\end{equation}
\\
where $\eta_0$ indicates the viscosity at zero shear rate, $\eta_{\infty}$ is the viscosity at infinite shear rate, $\dot\gamma_{c}$ is the critical shear rate for the onset of shear-thinning, $a$ is a dimensionless parameter controlling the smoothness of the transition between the low shear rate Newtonian plateau and the shear-thinning regime, and $n$ is the ``power law index'' in the shear-thinning regime \citep{Yasuda_2006}. The values of all those fitting parameters are listed in Table~\ref{tab:carreau}, along with the solvent-to-total viscosity ratio $\beta = \eta_s/\eta_0$. As is clear from the contrasting flow curves shown in Fig.~\ref{fgr:Rheology}(a), along with the contrasting values of $n$ and $\beta$ shown in Table~\ref{tab:carreau}, the two polymeric test solutions have significantly different shear-thinning characteristics. Hereafter, we refer to the $c =1$~wt\% PEO solution as ``highly shear-thinning'' (HS), and the $c =0.1$~wt\% PEO solution as ``weakly shear-thinning'' (WS). Note that, apart from the HS and WS test solutions described above, additional viscosity measurements made over a range of polymer concentrations $0.02 \leq c \leq 1$~wt\% enable estimation of the polymer overlap concentration $c^* \approx 0.05$~wt\% (see Fig.~\ref{fgr:Hspvsc}). Therefore, the HS ($c/c^* \approx 20$) and WS ($c/c^* \approx 2$) test solutions are both expected to be in the semidilute concentration regime, with the HS solution likely above the entanglement concentration \citep{Colby2010}.

\begin{table}
\begin{center}
\caption{\label{tab:carreau} Rheological parameters of the two poly(ethylene oxide) solutions tested in this study.}
%\begin{ruledtabular}
\begin{tabular}{cccccccc}
$c$ [wt\%] & $\eta_0$ [Pa s] & $\eta_{\infty}$ [mPa s] & $\dot\gamma_{c}$ [s$^{-1}$] & $n$ & $a$ & $\beta$ & $\tau_{EC}$ [s] \\
%\hline
0.1 & 0.018 & 3.0 & 6.42& 0.70 & 0.5 & 0.10 & 0.15\\
1.0 & 11.8 & 60 & 0.37 & 0.27 & 0.9 & 0.00015 & 0.38
\\
\\
\end{tabular}
%\end{ruledtabular}
 \end{center}
\end{table}
Fig.~\ref{fgr:Rheology}(b) shows the shear rate dependency of the first normal stress difference $N_1$ of each test fluid, measured during the same steady shear flow measurements as the viscosity flow curves shown in Fig.~\ref{fgr:Rheology}(a). For higher shear rates in the measurement range, both fluids exhibit measurable non-linear viscoelasticity (i.e., above the sensitivity limit of the normal force transducer).

\subsubsection{\label{sec:extensionrheo}Extensional Rheology}

Extensional rheological measurement is carried out at $T$ = 25 $^{\circ}$C using a capillary breakup extensional rheometer (CaBER 1, Thermo Haake). For this measurement, the fluid sample is loaded between the coaxial top and bottom plates of diameter $D_0 = 6$~mm with their initial separation set at 1~mm. The plates are then rapidly separated at a constant rate of $0.1~\text{m s}^{-1}$ until they reach their final separation of 10~mm. When the final position is reached, a laser micrometer positioned at the midpoint between the plates measures the diameter as a function of time $D(t)$ of the fluid filament as it thins towards pinchoff. The resulting time evolution of $D$ for both test solutions is shown in Fig.~\ref{fgr:Rheology}(c). Both fluids show clear elastocapillary (EC) thinning regimes during which the filament diameter decays exponentially in time, as is typical for viscoelastic fluids. The duration of this EC regime increases with polymer concentration, reflecting the greater elasticity and retarded capillary breakup of the more concentrated solution. The characteristic EC thinning time $\tau_{EC}$ is extracted from the slope in the EC regime according to $D(t) \sim \exp(-t/3\tau_{EC})$ \citep{Entov1997,Anna2001}. The resulting values of $\tau_{EC}$ are listed in Table~\ref{tab:carreau}. We note that $\tau_{EC}$ extracted from capillary thinning measurements is frequently referred to in the literature as being the ``relaxation time'' (or sometimes the ``extensional relaxation time''). However, due to recent work casting doubt over this description (e.g., \citet{Aisling2024,Gaillard2024,Calabrese2024,Calabrese2025}), we prefer here to call $\tau_{EC}$ the ``EC thinning time''. Despite this, we believe that the sensitivity of CaBER to elastic stress means that $1/\tau_{EC}$ should provide a relevant deformation rate beyond which elastic effects dominate viscous effects in extensional flows \citep{Calabrese2024,Calabrese2025}. 
Thus, $\tau_{EC}$ should provide a useful timescale by which to define the Weissenberg number (as discussed next). 

\subsection{Flow control and dimensionless numbers}\label{sec:dimension}

The flow of the test solutions through the microchannels is driven at controlled volumetric flow rates $Q$ using two Nemesys low-pressure syringe pumps (Cetoni GmbH). Both the inlet and the outlet of each channel are connected to a syringe by flexible silicone tubing. One of the pumps injects fluid into the channel, and the other pump withdraws the fluid from the outlet at the same volumetric flow rate. To parametrize the imposed flow conditions at different volumetric flow rates and quantify the effects of inertia and fluid elasticity, some key dimensionless numbers are introduced.

The Reynolds number Re quantifies the relative strength of inertial to viscous forces in the flow. In the system of this study, we consider two different Reynolds numbers. The first is based on the channel dimensions and is defined as:

 \begin{equation}
 \text{Re}_{ch} = \frac{\rho UD_h}{\eta(\dot\gamma_{w})},
\label{Re_ch}
 \end{equation}

where the density $\rho$ of the test fluids is considered equal to that of the solvent (i.e., 1290~kg m$^{-3}$), $U = Q/(WH)$ is the average flow velocity over the channel cross-section and $D_h = 2WH/(W+H) \approx 0.182$~mm is the hydraulic diameter of the straight rectangular cross-section microchannel. Here, $\eta(\dot\gamma_{w})$ is the shear rate-dependent viscosity obtained from the fits of the CY model evaluated at a wall shear rate of $\dot\gamma_{w} = 6U/H$. In most of the present experiments $\text{Re}_{ch} < 1$, but for the less viscous WS fluid at the higher end of the tested range of flow rates, a maximum value of $\text{Re}_{ch} \approx 5$ is reached. 

Another definition of the Reynolds number is based on the dimension $d$ of the cantilevers within the channel:

 \begin{equation}
 \text{Re} = \frac{\rho Ud}{\eta(\dot{\gamma})},
\label{Re}
 \end{equation}

where in this case $\eta(\dot{\gamma})$ is the shear rate-dependent viscosity obtained from the fits of the CY model evaluated at $\dot\gamma = U/d$. By this definition, the maximum Reynolds number in any of the present experiments is found to be $\text{Re} \approx 0.25$. Based on the Reynolds number estimates provided by Eqs.~\ref{Re_ch}~and ~\ref{Re}, we conclude that inertial effects on the flow in the vicinity of the cantilevers can be largely ignored.

The Weissenberg number, $\mathrm{Wi}$, represents the relative strength of elastic to viscous forces in the flow and is defined here as:

 \begin{equation}
 \label{tab:Wi}
 \text{Wi} = \tau_{EC} \dot{\gamma},
 \end{equation}

 where $\tau_{EC}$ is the characteristic EC thinning time of the fluid (given in Table \ref{tab:carreau}), and $\dot\gamma= U/d$ is the nominal shear rate near the cantilever. Although $\tau_{EC}$ is not interpreted here as a unique linear viscoelastic relaxation time, it provides an experimentally accessible elastic timescale under nonlinear extensional deformation and is therefore physically relevant to polymer stretching near the upstream and downstream stagnation regions of the cantilevers. We also consider an alternative definition for the Weissenberg number based on the first normal stress $N_1$ and shear stress $\sigma$ measured by rotational rheometry, namely $\text{Wi}_s = N_1/2\sigma$ \citep{Poole2012}, which can account for potential shear-thinning of the fluid relaxation time. 
 However, as shown by Fig.~\ref{fgr:N1vsWi}, over the range of shear rates where $N_1$ is measurable, there is, in fact, a one-to-one correspondence between $\mathrm{Wi}$ and $\text{Wi}_s$. Since the range of $N_1$ measurement is also quite narrow, and we prefer not to extrapolate the data, we select $\mathrm{Wi}$ given by Eq.~\ref{tab:Wi} as the primary control parameter to be used throughout the present flow experiments. 
 
 The elasticity number $\mathrm{El}$ quantifies the relative strength of elastic to inertial forces in the flow and is defined here by $\mathrm{Wi}$ and $\mathrm{Re}$:

\begin{equation}
 \label{tab:El}
 \text{El} = \text{Wi} / \text{Re},
 \end{equation}

In this experiment, the elasticity number spans $1 \times 10^{4} \lesssim \mathrm{El} \lesssim 2 \times 10^{4}$ for the WS fluid, and $3 \times 10^{5} \lesssim \mathrm{El} \lesssim 3 \times 10^{6}$ for the HS fluid, ensuring that elastic effects dominate for any applied flow rate such that $\text{Wi} \gtrsim 1$.

\subsection{Micro-particle image velocimetry}
 
The flow fields around the cantilever arrays are characterized using micro-particle image velocimetry ($\upmu$-PIV, TSI Inc.) \citep{Meinhart2000,Wereley2005,Wereley2010}. The test solutions are seeded with a low concentration ($c_{p}\approx 0.04$~wt\%) of red fluorescent polystyrene particles (Thermo Fisher Scientific Inc., diameter $0.4~\upmu$m, excitation/emission wavelength 542/612~nm). The $x$-$y$ plane of interest in the microchannel at $z = 0$  (i.e., the mid-height of the microfluidic device, Fig.~\ref{fgr_channel}(a)) is brought into focus on the imaging stage of an inverted microscope (Nikon Eclipse Ti) equipped with a Nikon PlanFluor objective lens ($10\times$ magnification, numerical aperture $NA = 0.3$). In this experiment, particle fluorescence is excited with a dual-pulse laser at a wavelength of 527~nm. The pulses are separated by the time duration $\Delta t$. A high-speed imaging sensor (Phantom MIRO) operating in frame-straddling mode enables the capture of image pairs synchronized with the laser pulses. Note that the measurement depth under these imaging conditions is $\delta_z \approx 26.5~\upmu$m \citep{Meinhart2000}, or approximately one-quarter of the channel and cantilever height. At each imposed flow rate, the time $\Delta t$ is set such that the mean particle displacement in the streamwise direction between image pairs is approximately 6 pixels. Up to 1000 pairs of particle images are captured and processed using commercial PIV software (Davis 10.2.1, LaVision GmbH) with a multi-pass iterative cross-correlation algorithm with window deformation starting from a rectangular grid of $204.8 \times 102.4~\upmu$m ($x \times y$) and refining to a final interrogation window of $12.8 \times 6.4~\upmu$m with 50~\% overlap. The rectangular window is selected to account for the anisotropic nature of the flow field within the cantilever array, where the streamwise velocity component is dominant over the transverse component. The longer window dimension is aligned with the primary flow along $x$ in order to capture sufficient particle pairs for cross-correlation, while the shorter dimension in the transverse $y$ direction preserves spatial resolution in the regions between adjacent cantilevers. Since for the majority of imposed flow rates the flows around the cantilever arrays are in any case time-steady, we present time-averaged velocity fields in order to illustrate the dominant flow state observed at each $\text{Wi}$. All subsequent post-processing and image analysis is performed in MATLAB.

 \subsection{Cantilever tracking}

The flow-induced motion of each cantilever is imaged using an inverted microscope (Nikon Eclipse Ti) equipped with a Nikon PlanFluor objective lens ($10\times$ magnification, $NA = 0.3$). An additional $1.5 \times$ magnification is provided by an extra tube lens inserted using the intermediate magnification dial of the microscope. The focal plane is set at the tips of the cantilevers ($z=46~\upmu$m) and images are recorded for 20~s at between 100 and 1600 frames-per-second (depending on the imposed flow rate) using a high-speed imaging sensor (Phantom MIRO). The tip position of each cantilever is detected using the built-in \textit{imfindcircles} function from MATLAB, from which the center coordinates $(x(t), y(t))$ are extracted. The resting position of each cantilever $(x_0,y_0)$, recorded under no-flow conditions, is subtracted from its position in each frame recorded under flow. The most probable deflection position of each cantilever over the 20~s recording period $(\Delta x,\Delta y)$ is determined from the peak of the probability density distribution of $(x(t)-x_0,y(t)-y_0)$ computed in MATLAB.

\begin{figure*}
%\centering
  \includegraphics[scale=0.245]{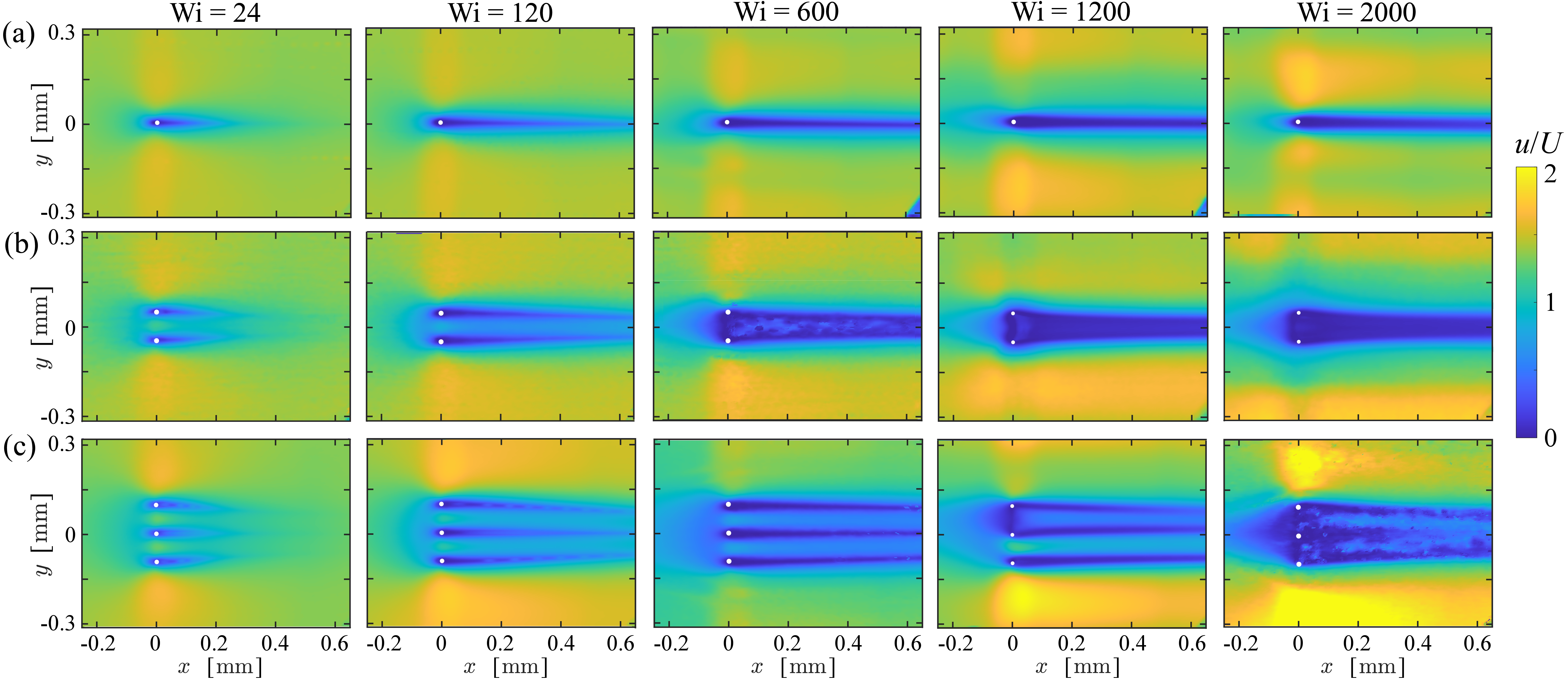}
  \caption{Time averaged velocity fields around the cantilever arrays with weakly shear-thinning (WS) fluid: (a) single, (b) double, (c) triple cantilever configurations, which are aligned side by side in the middle of the microchannel. The color contours indicate the normalized magnitude of the streamwise velocity $u/U$. Flow is from left to right, and the values of the Weissenberg number Wi are given at the top of each column.}
  \label{fgr_boger}
\end{figure*}

\begin{figure*}
\centering
  \includegraphics[scale=0.245]{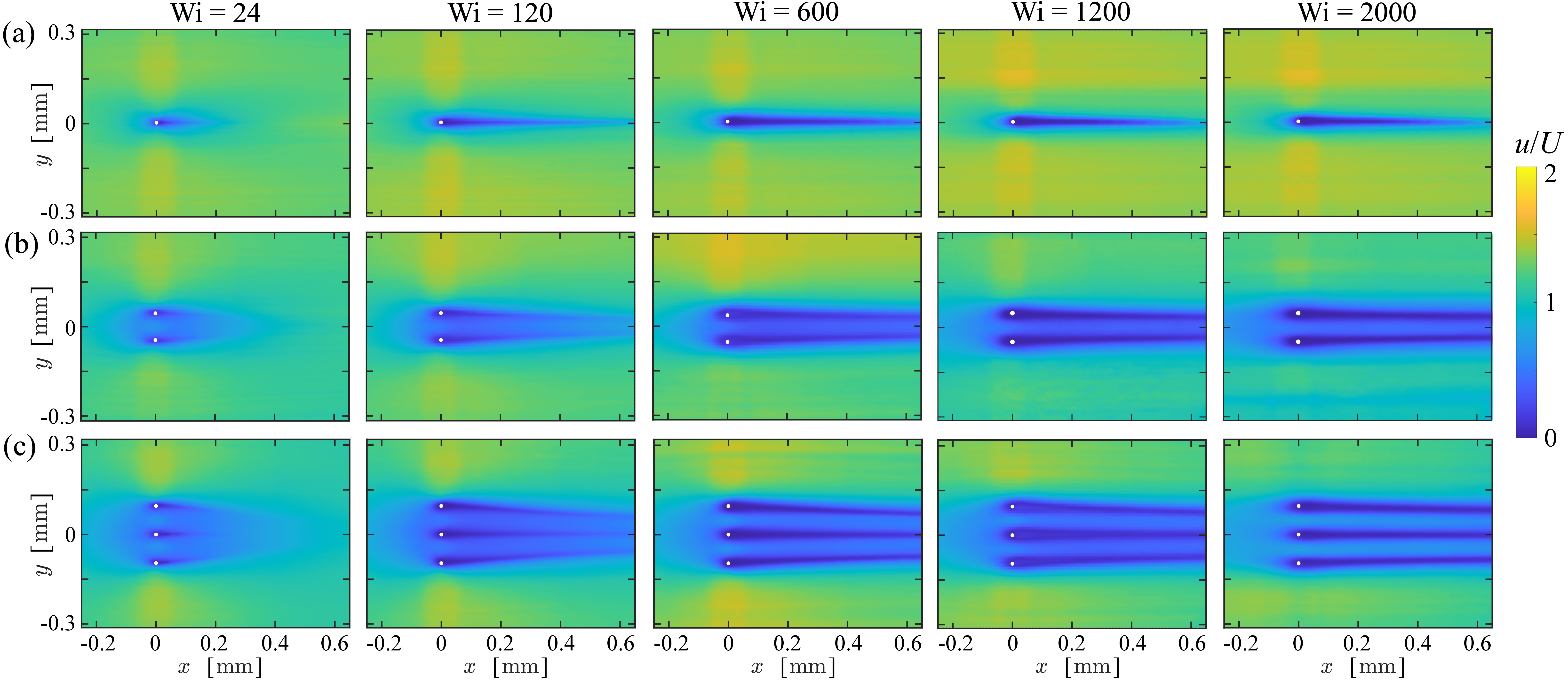}
  \caption{Time averaged velocity fields around the cantilever arrays with highly shear-thinning (HS) fluid: (a) single, (b) double, (c) triple cantilever configurations, which are aligned side by side in the middle of the microchannel. The color contours indicate the normalized magnitude of the streamwise velocity $u/U$. Flow is from left to right, and the values of the Weissenberg number Wi are given at the top of each column.}
  \label{fgr:shear-thinning}
\end{figure*}

\section{Results and discussion}

\subsection{Flow fields around cantilever arrays}
\label{sec:flowfield}

Representative time-averaged velocity fields for the WS ($c = 0.1$~wt\%) and HS ($c = 1$~wt\%) PEO solutions are shown in Fig.~\ref{fgr_boger} and Fig.~\ref{fgr:shear-thinning}, respectively, over the same ranges of imposed Weissenberg number Wi. Each set of images illustrates the variety of flow states observed across single, double, and triple cantilever configurations. At the top of each column, the dimensionless number Wi is displayed to quantify the degree of fluid elasticity for each flow condition, as defined in Sec.~\ref{sec:dimension}. Note that the flow propagates from the left to the right, and for comparison Fig.~\ref{fgr:newtonian_table} shows the velocity field around the cantilever arrays for a Newtonian reference fluid at $\text{Re} \approx 0.25$ (equivalent to the maximum Reynolds number reached in the tests with the viscoelastic fluids). From the velocity fields for Newtonian fluid, the absence of vortex shedding around the cantilevers confirms that inertial effects can be neglected throughout the experiments \citep{Peng2020}. Compared to the velocity field for Newtonian fluid shown in Fig.~\ref{fgr:newtonian_table}, for both polymer solutions (Figs.~\ref{fgr_boger} and \ref{fgr:shear-thinning}) elongated wakes of low flow velocity are observed downstream of each cantilever, and develop with increasing Wi. Such elongated wakes are typical for viscoelastic flows around objects, and are associated with polymer stretching and the elastic response of fluids to the strong velocity gradients near the upstream and downstream stagnation points \citep{Chilcott1988,Mckinley1993,Solomon1996,Nolan2016,Haward2018,Varchanis2020}.

For the WS fluid, in the case of the single cantilever (Fig.~\ref{fgr_boger}(a)), the elastic wake grows in spatial extent as Wi is increased, while remaining laterally symmetric about $y = 0$ throughout the range of Wi investigated. However, the flow fields around arrays containing more than one cantilever show a transition in form as Wi is increased. For the double cantilever configuration (Fig.~\ref{fgr_boger}(b)), at lower Wi (e.g., $\text{Wi}=24$ and $\text{Wi}=120$), the elastic wakes around the two cantilevers are separated and distinct. However, as Wi is increased beyond a critical value $\text{Wi}^{*}$, the elastic wakes trailing each cantilever merge with one another, as shown in Fig.~\ref{fgr_boger}(b) for $\text{Wi}=600$ and greater. The merged wake region, where the normalized streamwise velocity approaches zero, effectively shields the inter-cantilever gap from the incoming flow. As a result, the upstream flow decelerates as it approaches the two cantilevers and is forced to diverge around the outer sides of the cantilevers. The resulting appearance is of the two cylinders behaving as a single larger obstacle to the incoming flow. Similar ``diverging flow states'' were also observed around side-by-side rigidly-fixed cylinders in previous experimental and numerical flow studies \citep{Hopkins2021, Khan2021, Mokhtari2022}.

Qualitatively similar diverging flow states are observed in the triple cantilever configuration (Fig.~\ref{fgr_boger}(c)). In this case, it is possible for the wakes of two adjacent cantilevers to merge while the remaining third wake remains isolated (as shown for $\text{Wi}=1200$, Fig.~\ref{fgr_boger}(c)), resulting in a ``partially diverging state''. In this partially diverging state, the location of the merged wake pair is stochastic, the pair of wakes that merges varies randomly each time the flow is stopped and restarted at the same Wi.  At higher Wi (e.g., $\text{Wi}=2000$, Fig.~\ref{fgr_boger}(c)), the wakes of all three cantilevers unify into a single large low-velocity region, with all three cantilevers now effectively acting as a single collective obstacle to the incoming flow and confining high streamwise velocities to the regions flanking the array. Similar flow behavior was also reported in a numerical study on viscoelastic flows past three side-by-side but rigidly-fixed cylinders \citep{Khan2022}. Notably, in both the double and triple configurations, the onset of the diverging flow state is accompanied by the appearance of a markedly low-velocity region extending upstream of the pair, or triplet, of cantilevers with merged wakes (see e.g., $\text{Wi} \geq 600$ in Fig.~\ref{fgr_boger}(b), and $\text{Wi} \geq 1200$ in Fig.~\ref{fgr_boger}(c)). 

Flow velocimetry performed with the HS PEO solution around the single cantilever arrays (Fig.~\ref{fgr:shear-thinning}(a)) is qualitatively similar to that seen with the WS fluid (Fig.~\ref{fgr_boger}(a)), with the growth of an increasingly long downstream wake, that remains aligned on the $y=0$ axis as the Weissenberg number is increased up to very high values $\text{Wi}=2000$. This result is perhaps surprising since previous experimental and theoretical studies on viscoelastic flows around single (rigidly-fixed) cylinders have shown that stronger shear-thinning promotes asymmetry of the downstream wake and preferential flow around one side of the cylinder \citep{Haward2019,Haward2020,Varchanis2020,Khan2021,Haward2021,Spyridakis2024}. Experimental work on shear-thinning viscoelastic flows past cantilevered cylinders have also shown similar asymmetries to those observed with fixed cylinders (see e.g., \citet{Hopkins2020}). The difference between the current and prior works may be explained by the contrasting flow geometries used. While prior works have involved 2D or quasi 2D flow with high aspect ratio $A_R= H/W > 1$, in the present work we use a channel with a significantly lower value of $A_R = 0.1$, meaning the flow should be parabolic through the channel depth $z$ as opposed to across the channel width through $y$. A contrasting blockage ratio may also be important. While prior works typically considered relatively large obstacles, with cylinders blocking the channels by a factor $B_R = d/W \geq 0.1$, the much lower value of $B_R = 0.008$ in the present study means the fluid velocity increases only mildly around the sides of the cylinder, generating smaller differences in local shear rates, and consequently reducing the contribution of shear-thinning to the in-plane flow behavior.  

Although the flow fields around the single cantilever are similar for the HS and WS fluids, the velocimetry of the HS fluid around the double and triple cantilever arrays (Fig.~\ref{fgr:shear-thinning}(b,c)) reveals a qualitatively different behavior from that seen with the WS fluid at equivalent Wi values. For the HS fluid, no merging of neighboring wakes is observed even up to the maximum imposed $\text{Wi}=2000$. We do not attribute the different behavior to the possibility of shear-thinning relaxation times leading to different effective Weissenberg numbers between the two fluids: as indicated by Fig.~\ref{fgr:N1vsWi}, equivalence of $\mathrm{Wi}$ between the two fluids implies equivalence of $\text{Wi}_s$ also. This suggests that the difference in either the magnitude of the first normal stress difference $N_1$ (Fig.~\ref{fgr:Rheology}(b)) or in the degree of shear-thinning between the fluids (Fig.~\ref{fgr:Rheology}(a)) may play the critical role in governing the onset of the wake merger and flow divergence in this geometric system. Here, the collapse shown in Fig.~\ref{fgr:N1vsWi} further suggests that the elastic instability criterion of Pakdel and McKinley \citep{Mckinley1996,Pakdel1996} does not account for the distinct stability behavior exhibited by the two viscoelastic test solutions. The criterion of Pakdel and McKinley predicts the onset of a purely elastic instability when the parameter $M^2 = \tau U/\mathcal{R} \cdot N_1/\sigma$ exceeds a critical value, where $\mathcal{R} = d / 2$ is a characteristic streamline radius of curvature. This can effectively be rewritten as $M^2 \approx 4\text{Wi} \cdot \text{Wi}_s$. Hence, for a given value of the Weissenberg number the $M^2$ parameter should be approximately the same for both the WS and the HS fluid, yet the stability behavior is clearly different, indicating that the elastic instability criterion alone is insufficient to distinguish the onset of instability in the present system.

\subsection{Divergent flow}
 \label{sec:flowasymmetry}
 
\begin{figure}
\centering
  \includegraphics[scale=0.4]{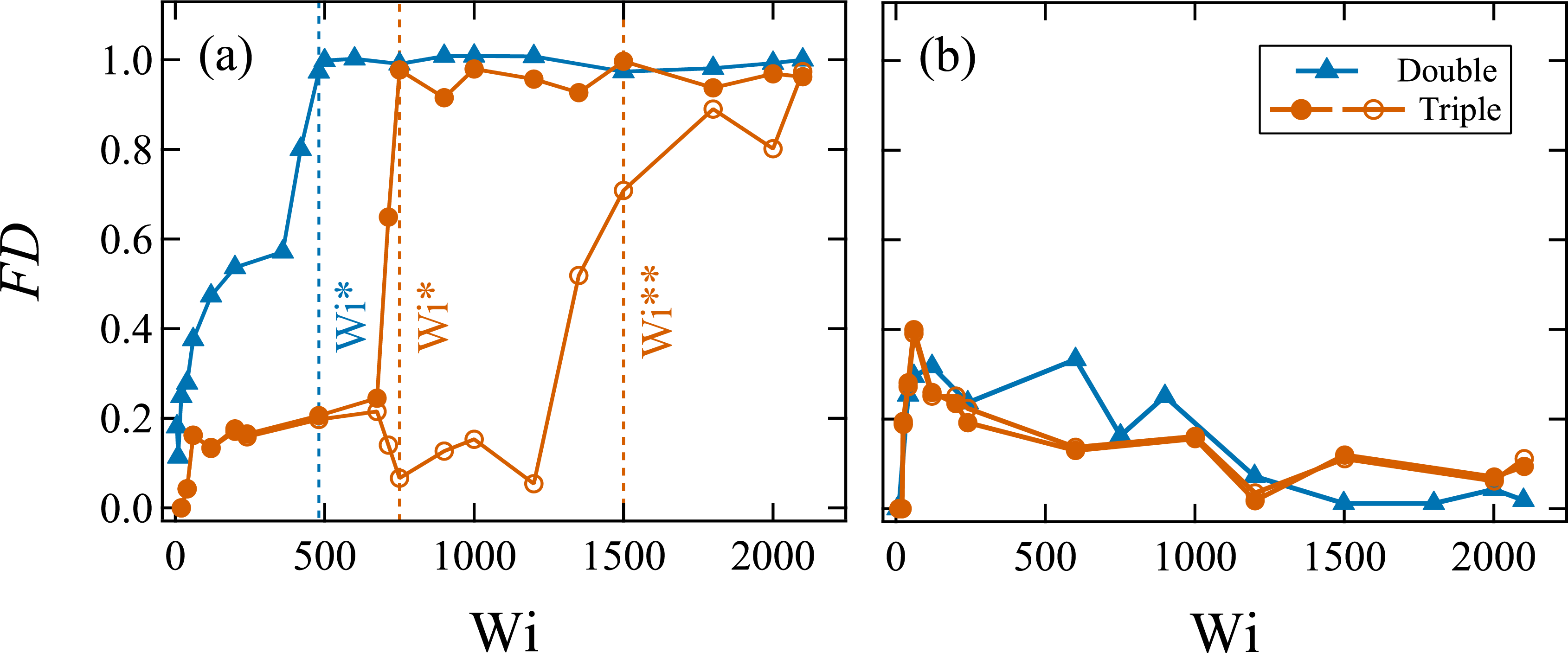}
  \caption{Quantification of the divergent-flow transition. Flow divergence ratio $FD$ (see  Eq.~\ref{eq:divergent}) as a function of the Weissenberg number $\mathrm{Wi}$  for the double and triple cantilever arrays in (a) weakly shear-thinning (WS) fluid, (b) highly shear-thinning (HS) fluid.}
  \label{fgr:asymmetryratio}
\end{figure}

From the velocity fields measured by $\upmu$-PIV over a range of Wi, we assess the degree of divergence for the flow at each inter-cantilever gap by using the following flow divergence ratio $FD$: 

 \begin{equation}
 \label{eq:divergent}
 FD = \frac{u_g^*(\text{Wi}=6)-u_g^*(\text{Wi})}{u_g^*(\text{Wi}=6)},
 \end{equation}
 \\
where $u_g^* = u_g/U$, and $u_g$ is the streamwise flow velocity in the gap of interest between the cantilevers. The Weissenberg-dependent value $u_g^*(\text{Wi})$ is compared to a low-Wi reference condition $u_g^*(\text{Wi}=6)$ ($\text{Wi}=6$ being the lowest Wi investigated). Accordingly, a value of $FD = 0$ corresponds to no change from the low-Wi reference state, while increasing $FD>0$ indicates an increasing reduction in local streamwise velocity, reflecting the development of divergent flow. A value of $FD \rightarrow 1$ means that the flow velocity at the intercylinder gap is close to zero.

The flow divergence ratio $FD$ (Eq.~\ref{eq:divergent}) is plotted as a function of Wi for the two polymeric test solutions in Fig.~\ref{fgr:asymmetryratio}. For the WS fluid (Fig.~\ref{fgr:asymmetryratio}(a)), $FD$ increases gradually with Wi below the critical Weissenberg number $\text{Wi}^{*}$, consistent with the progressive elongation of the elastic wake trailing each cantilever. For the double cantilever configuration, at $\text{Wi}^{*} \approx 480$ the value of $FD$ increases abruptly to the maximum value of $FD \approx 1$, corresponding to a near-complete suppression of the flow through the intercantilever gap. Subsequently, with further increasing Wi, the value of $FD$ remains at or close to its maximum value. This sharp transition marks the onset of wake merger between the neighboring cantilevers, as observed in the velocity fields of Fig.~\ref{fgr_boger}(b) at $\text{Wi} \geq 600$. For the triple cantilever configuration, for $\text{Wi} \lesssim 750$ the wakes behind the cantilevers remain separated and the flow around the cantilevers is steady in time. Subsequently, as Wi is increased, there are two transitions across the two gaps. The first (shown by solid circles in Fig.~\ref{fgr:asymmetryratio}) occurs abruptly at $\text{Wi}^{*} \approx 750$ and corresponds to the steady merger of two out of the three downstream wakes (as illustrated by Fig.~\ref{fgr_boger}(c) at $\text{Wi}=1200$). 
 A second transition (shown by open circles in Fig.~\ref{fgr:asymmetryratio}) occurs above a second critical Weissenberg number ($\text{Wi}^{**} \approx 1500$), corresponding to the development of the fully unified wake state observed at $\text{Wi}=2000$ in Fig.~\ref{fgr_boger}(c). Note that this second transition occurs over a range of $1200 \leq \text{Wi} \leq 1800$, over which we observe the time-dependent behavior characterized by the transient switching between the existence of a partially and a fully unified wake. Fig.~\ref{fgr:I_error} shows the divergent ratio $FD$ as a function of Wi, including the error bounds. The error bounds for the triple cantilever array are notably larger than the results from the double cantilever configuration and their magnitude increases gradually for $\text{Wi} \geq 1200$, indicating how the unified wake state gets more dominant over the experimental observation time as Wi is increased. 

The higher onset threshold $\text{Wi}^{*}$ observed for the triple cantilever array may be associated with the presence of two geometrically equivalent intercantilever gaps. In the double configuration, wake interaction is confined to a single gap, so the elastic wake merger occurs at a relatively low critical value, $\text{Wi}^{*} \approx 480$. In the triple configuration, by contrast, either of the two equivalent gaps can support wake interaction, and the system initially selects one pair of neighboring wakes to merge, leading to a partially divergent state at $\text{Wi}^{*} \approx 750$. The stochastic selection of the merged pair, which varies between repeated flow start-ups at the same Wi, further indicates that the two gaps are dynamically equivalent and that small fluctuations in the elastic stress field determine which local wake-merger event occurs first. A second transition at higher Wi then corresponds to the development of a fully unified wake across all three cantilevers.

For the HS fluid (Fig.~\ref{fgr:asymmetryratio}(b)), the trend is markedly different from that observed for the WS fluid. The value of $FD$ increases slightly to a value of $FD \approx 0.4$ at lower values of Wi, reflecting the growth of elastic wakes. However, unlike the WS fluid, no abrupt transition is observed at higher Wi; instead, the value of $FD$ remains at a low value across the entire range of Wi investigated. This analysis effectively quantifies the evolution with Wi of the wake structures shown in Fig.~\ref{fgr_boger} and Fig.~\ref{fgr:shear-thinning} for the WS and HS fluid, respectively.

From the flow velocimetry performed in all cantilever configurations and with both polymeric fluids (see Figs.~\ref{fgr_boger} and~\ref{fgr:shear-thinning}), there is often a slight velocity contrast between the flow velocity on either side of the cantilever arrays. Since prior theoretical and experimental works have reported flow asymmetries in related geometries \citep{Dey2018,Haward2019,Hopkins2020,Haward2020,Varchanis2020,Khan2021,Haward2021,Spyridakis2024}, this velocity contrast is also examined here by parameterizing a flow asymmetry ratio $FA$:
 
 \begin{equation}
 \label{eq:asymmetry}
 FA = \frac{|u_{side,1}-u_{side,2}|}{u_{side,1}+u_{side,2}},
 \end{equation}
where $u_{side,i}$ is the averaged streamwise velocity measured on either side of the cantilever array. A value of $FA = 0$ therefore corresponds to a perfectly symmetric flow, with no difference in streamwise velocity around the sides of the array. A value of $FA \rightarrow 1$ indicates that the streamwise velocity on one side of the array approaches zero, representing a fully asymmetric flow state in which flow is predominantly directed around one side of the array only.

As shown in Fig.~\ref{fgr:I_star}, for both the WS and the HS fluid the flow asymmetry parameter is almost negligible compared to the flow divergence parameter. The flow around the side of the arrays remains essentially symmetric over the entire range of Weissenberg number, including throughout the onset and development of merging wake structures.

\subsection{Cantilever deflection in viscoelastic flow}

\begin{figure*}
\centering
  \includegraphics[scale=0.245]{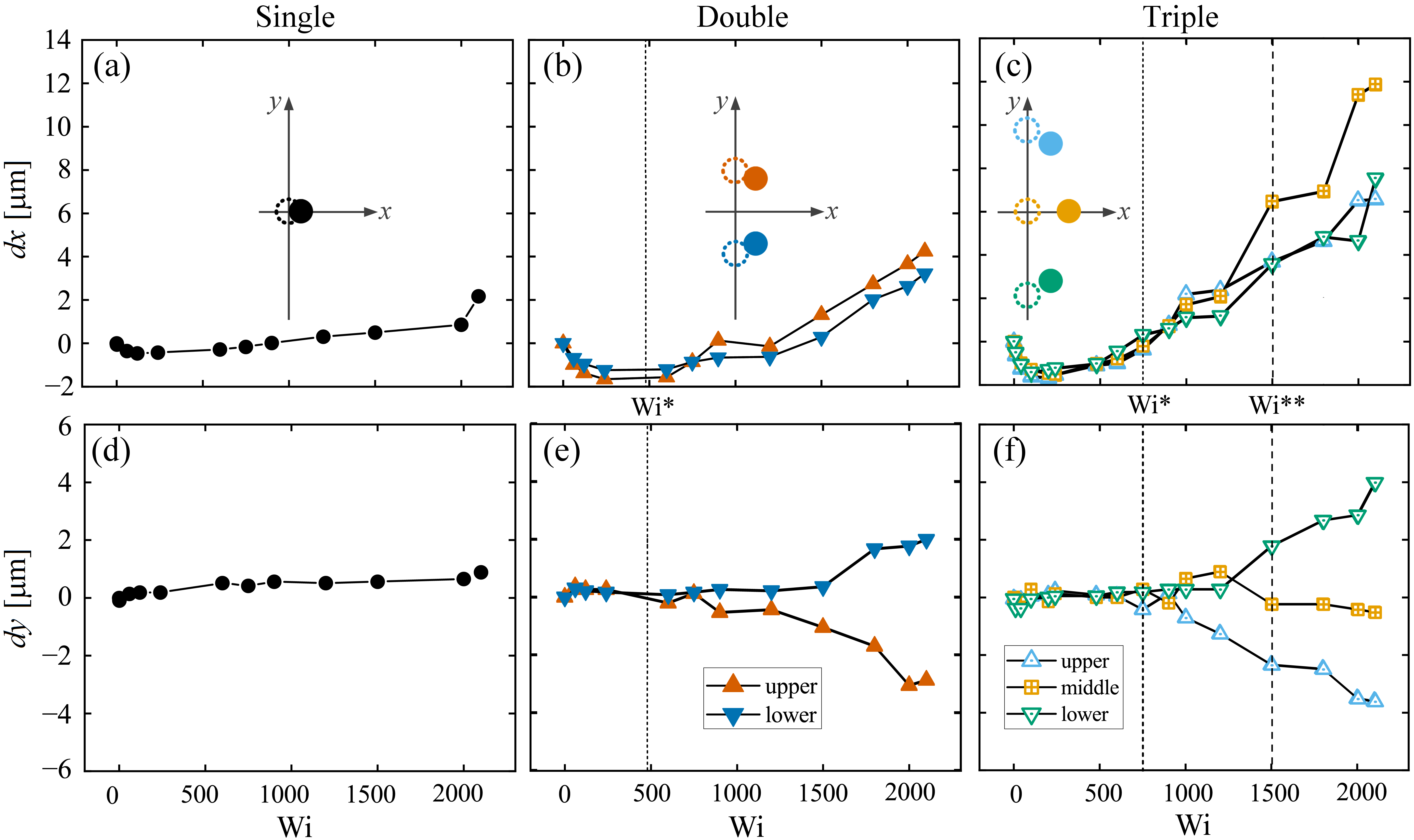}
  \caption{Cantilever deflection in the weakly shear-thinning (WS) fluid relative to a zero viscosity-matched Newtonian fluid. The plotted quantities are difference between the absolute deflection of the flexible cantilevers measured in the WS fluid ($\eta_0 = 18$~mPa~s) and in a Newtonian fluid of viscosity $\eta = 18$~mPa~s. (a)-(c) Streamwise deflection difference $dx = \Delta x_{WS} - \Delta x_{Newt}$, and (d)-(f) spanwise deflection difference $dy = \Delta y_{WS} - \Delta y_{Newt}$ for different cantilever configurations: single (left column), double (middle column), and triple (right column). Insets in (a)--(c) schematically show the tip displacement of the cantilevers, with open dashed circles represent their initial positions at $\text{Wi} = 0$ and solid circles represent their positions at $\text{Wi} = 2000$.
}
  \label{fgr:pillar-relative_Newtonian}
\end{figure*}

\subsubsection{Comparison with Newtonian fluid}
\label{sec_pillardis_new}

We first examine the structural response of the cantilevers in a Newtonian fluid (a 70~wt\% glycerol and 30~wt\% water mixture), chosen to closely match the zero shear viscosity of the WS fluid ($\eta_{0} = 18$~mPa~s). The streamwise and spanwise cantilever deflection ($\Delta x$ and $\Delta y$) under Newtonian flow over a range of volumetric flow rates is shown in Fig.~\ref{fgr:Newtonian_pillardis} (note here the maximum Reynolds number is $\text{Re} \approx 0.1$). The streamwise deflection of the cantilever reaches a plateau as the volumetric flow rate is increased, consistent with the behavior qualitatively reported by \citet{Wexler__2013} and a similar trend is observed for all of the cantilevers in the various arrays: single, double and triple. As shown in Fig.~\ref{fgr:Newtonian_pillardis}, the spanwise deflection of all the cantilevers in all three arrays is close to $\Delta y = 0$.

In Fig.~\ref{fgr:pillar-relative_Newtonian}, we compare the absolute displacement of the cantilevers in the WS fluid against that of the Newtonian fluid with matched viscosity (shown in Fig.~\ref{fgr:Newtonian_pillardis}): $dx(Q) = \Delta x_{WS}(Q) - \Delta x_{Newt}(Q)$, $dy(Q) = \Delta y_{WS}(Q) - \Delta y_{Newt}(Q)$. For the single cantilever configuration, $dx$ decreases very slightly at low $\mathrm{Wi}$, before showing a gradual increase with increasing $\mathrm{Wi}$, indicating a slightly greater deflection in the WS fluid than seen in the Newtonian case (Fig.~\ref{fgr:pillar-relative_Newtonian}(a)). The initial low-Wi decrease in $dx$ is more obvious for the double (Fig.~\ref{fgr:pillar-relative_Newtonian}(b)) and triple (Fig.~\ref{fgr:pillar-relative_Newtonian}(c)) cantilever cases, as does the subsequent increase in $dx$ with increasing Wi, which appears to be nonlinear for $\text{Wi} \gtrsim 1000$. The initial small decrease in $dx$ at low Wi may be attributed to the slight shear-thinning of the WS fluid , which could be expected to reduce the drag force acting on each cantilever relative to the Newtonian case. We attribute the subsequent increase in $dx$ at higher Wi to the increase in extensional viscosity due to polymer stretching around the cantilevers (particularly in the downstream wake regions), which induces the enhancement of drag on the cantilevers, driving the nonlinear increase in $dx$ \citep{Francois2008}. We note how the streamwise deflection of the cantilevers increases with the number of cantilevers in the array, suggesting a cooperative interaction, presumably related to the merging of extensional wakes. 
For the triple cantilever array, we further note how $dx$ for the middle cantilever exhibits the greatest deflection of the three for Weissenberg numbers beyond $\text{Wi} \approx 1500$, as seen in Fig.~\ref{fgr:pillar-relative_Newtonian}(c), almost coinciding with the second flow transition at $\text{Wi}^{**}$, identified in Fig.~\ref{fgr:asymmetryratio}(a). This correlation suggests that the reduced flow permeability through the cantilever array, arising from the fully unified wake, drives the flow to accelerate around the sides of the cantilevers rather than between them, thereby enhancing the drag force on, and deflection of, the middle cantilever in particular. 

In the spanwise direction, the single cantilever shows no measurable difference from the Newtonian fluid, thus $dy$ remains close to zero across the entire range of Wi investigated (Fig.~\ref{fgr:pillar-relative_Newtonian}(d)). This simple streamwise deflection of the single cantilever is depicted in the inset to Fig.~\ref{fgr:pillar-relative_Newtonian}(a), where the dotted circle represents the position of the cantilever tip at Wi~=~0, and the filled circle represents the position at Wi~=~2000. However, the presence of neighboring cantilevers introduces a qualitatively different response. In the double cantilever configuration (Fig.~\ref{fgr:pillar-relative_Newtonian}(e)), as Wi is increased beyond the first transition at $\text{Wi}^* \approx 480$, the two cantilevers deflect towards each other, converging toward the center of the array as depicted in the insert to Fig.~\ref{fgr:pillar-relative_Newtonian}(b). In the triple cantilever configuration (Fig.~\ref{fgr:pillar-relative_Newtonian}(f)), the middle cantilever and the upper cantilever weakly deflect toward each other beyond the first transition at $\text{Wi}^* \approx 750$, consistent with the merging of two out of the three downstream wakes seen in Fig.~\ref{fgr_boger}(c) at Wi~=~1200. Subsequently, for $\text{Wi} \gtrsim 1500$, the middle cantilever returns toward its centerline position ($dy \approx 0$) while the upper and lower cantilevers deflect toward the center of the array (as depicted in the insert to Fig.~\ref{fgr:pillar-relative_Newtonian}(c)), consistent with the second transition to the fully unified wake state occurring at $\text{Wi}^{**} \approx 1500$.

Clearly, we see a strong correlation between the inward spanwise deflection of the flexible cantilevers, the macroscopic emergence of merged wakes downstream of the cantilevers, and the development of divergent flow around the cantilever arrays.

\subsubsection{Effect of fluid rheology}
\label{sec_pillardis_non}

\begin{figure*}
\centering
  \includegraphics[scale=0.35]{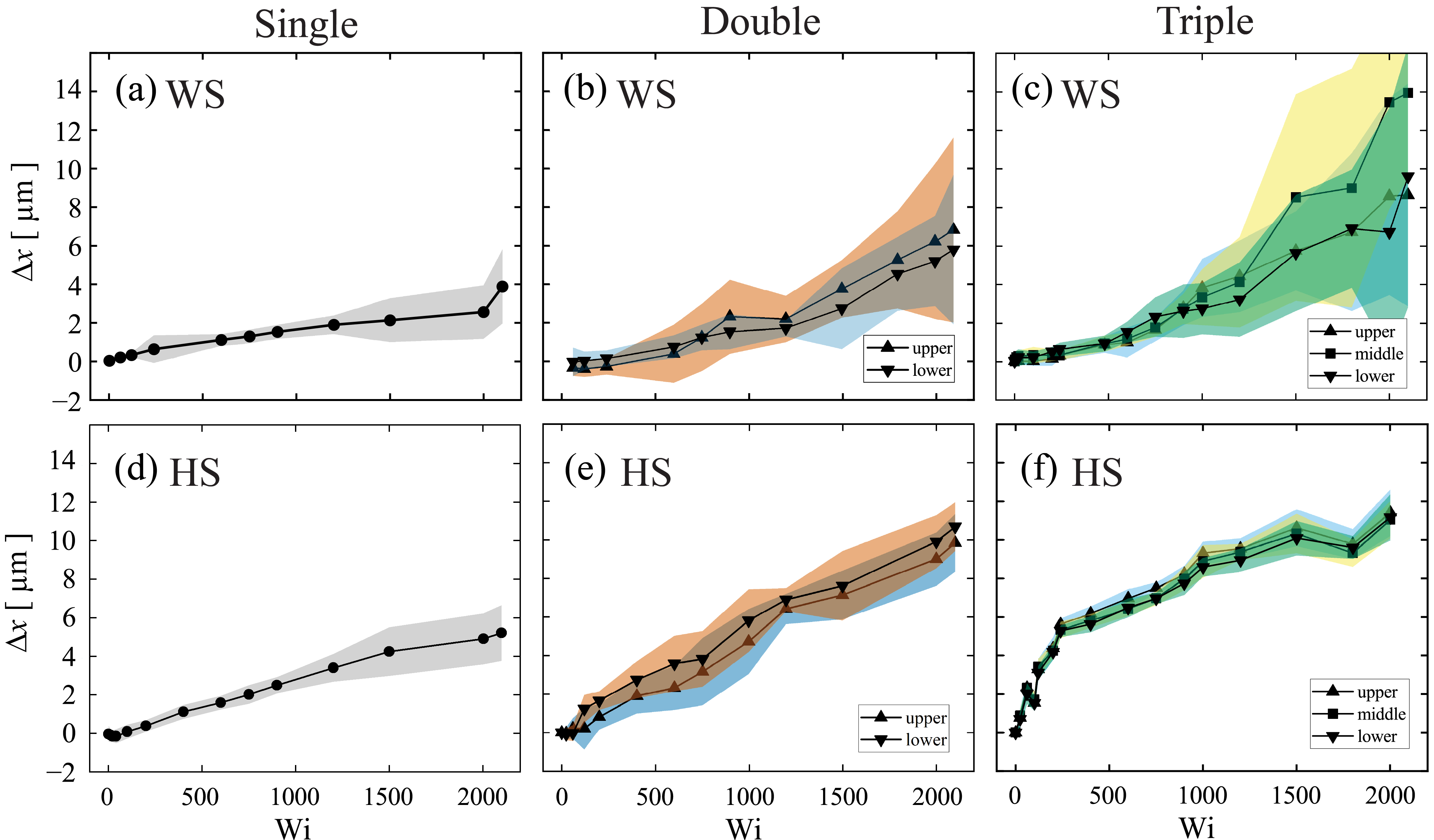}
  \caption{Absolute streamwise displacement $\Delta x$ of each cantilever as a function of the Weissenberg number $\mathrm{Wi}$ for weakly shear-thinning fluid (WS) (upper) and highly shear-thinning (HS) (lower) test solutions: (a), (d) single cantilever, (b), (e) double cantilever, (c), (f) triple cantilever.
}
  \label{fgr:nonNewtonians_dx}
\end{figure*}

\begin{figure*}
\centering
  \includegraphics[scale=0.35]{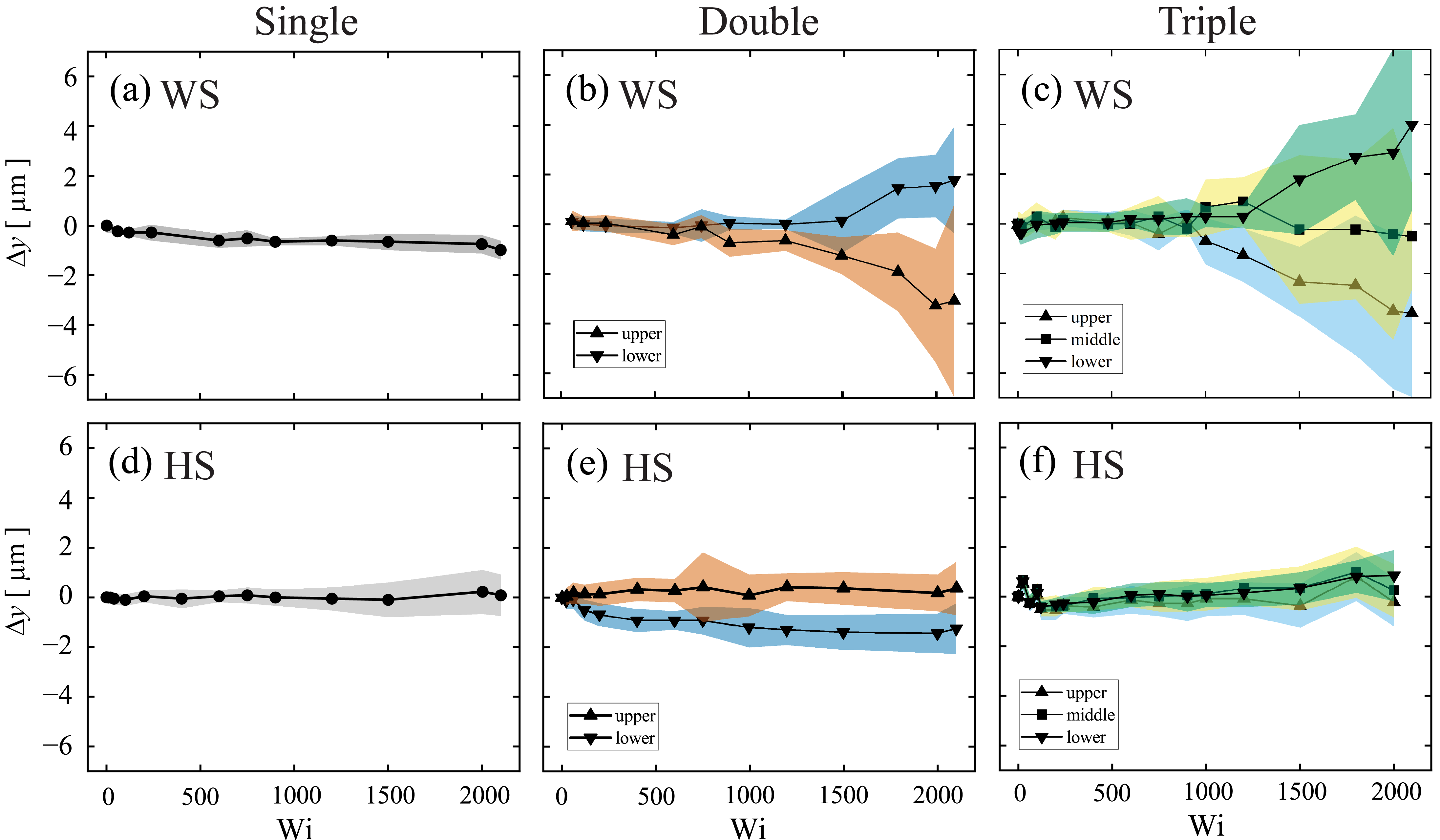}
  \caption{Absolute spanwise displacement $\Delta y$ of each cantilever as a function of the Weissenberg number $\mathrm{Wi}$ for weakly shear-thinning fluid (WS) (upper) and highly shear-thinning (HS) (lower) test solutions: (a), (d) single cantilever, (b), (e) double cantilever, (c), (f) triple cantilever.
}
  \label{fgr:nonNewtonians_dy}
\end{figure*}

We now compare the WS  ($c = 0.1$~wt\%) and HS ($c = 1.0$~wt\%) PEO solutions to examine how fluid rheology influences the wake-mediated collective instability identified above. The absolute values of the cantilever deflection in the streamwise direction $\Delta x$ and in the spanwise direction $\Delta y$ are plotted as a function of Wi in Fig.~\ref{fgr:nonNewtonians_dx} and Fig.~\ref{fgr:nonNewtonians_dy}, respectively.

For the WS fluid, the value of $\Delta x$ increases monotonically with increasing $\mathrm{Wi}$ across all cantilever configurations (Fig.~\ref{fgr:nonNewtonians_dx}(a,b,c)), with a general nonlinear upward curvature at higher Wi suggesting the dominance of extension thickening over shear-thinning effects. The temporal variability during the experiment is indicated by the shaded regions around the data points, which show the width of the probability density distribution at 10~\% of its peak value. This variability increases progressively with increasing $\text{Wi}$, indicating that the cantilever motion becomes increasingly unsteady.

For the HS fluid, the value of $\Delta x$ also increases monotonically with $\mathrm{Wi}$, exhibiting a much stronger Wi dependence than the WS fluid at low $\mathrm{Wi}$ due to the higher viscosity, but with a definite downwards curving trend suggestive of the continued influence of shear-thinning (as seen in Fig.~\ref{fgr:nonNewtonians_dx}(d,e,f)). In contrast to the WS fluid, the error bounds for the HS fluid are narrower, indicating that the cantilever motion is less time-dependent. Also notable in Fig.~\ref{fgr:nonNewtonians_dx}(f) is that for the HS fluid all three cantilevers deflect equally downstream, in contrast to the WS fluid for which, at higher $\mathrm{Wi}$, the central cantilever deflects significantly further than those at the edges.

The clearest distinction between the two fluids emerges in the spanwise deflection of the cantilevers, as shown in Fig.~\ref{fgr:nonNewtonians_dy}. For the WS fluid, $\Delta y$ grows significantly at high $\mathrm{Wi}$, with the cantilevers deflecting inward toward the center of the array, consistent with the merged-wake state identified in Sec.~\ref{sec_pillardis_new}. In particular, for the triple cantilever configuration in the WS fluid, the time-dependence of the cantilever displacement is pronounced in both $\Delta x$ and $\Delta y$, as evidenced by the large error bounds in Fig.~\ref{fgr:nonNewtonians_dx}(c) and Fig.~\ref{fgr:nonNewtonians_dy}(c). This is consistent with the time-variant flow state discussed in Sec.~\ref{sec:flowasymmetry}. 
For the HS fluid, the value of $\Delta y$ remains small with values in the range $\lvert{\Delta y}\lvert$$<$ 1.5~$\upmu$m for all the cantilever geometries, consistent with the symmetric velocity field maintained throughout the range of $\mathrm{Wi}$ investigated, as shown in Fig.~\ref{fgr:shear-thinning}. This result is completely different from previous studies involving shear-thinning viscoelastic flows around deformable cylinders, in which 2D motions were reported (see Sec.~\ref{sec:intro}) \citep{Dey2018,Hopkins2020}, and this discrepancy is also consistent with the observations made on the flow fields (Sec.~\ref{sec:flowfield}). 

The inward spanwise deflection of the cantilevers observed in the WS fluid could conceivably arise from higher shear rates along the outer edges of the cantilevers and the resulting effects of the first normal stress difference $N_1$. However,  $N_1$ is greater for the HS fluid than for the WS fluid at a given shear rate (see Fig.~\ref{fgr:Rheology}(b)), whereas the HS fluid exhibits neither substantial inward deflection nor wake merger. This suggests that the structural response cannot be explained by the magnitude of $N_1$ effects alone. The principal remaining rheological distinction between the two fluids is therefore their degree of shear-thinning, indicating that shear-thinning strongly influences whether neighbouring elastic wakes can merge and, consequently, whether the divergent-flow state and coordinated spanwise deflection associated with the collective instability can develop. 

\subsection{Physical interpretation of collective instability}

The results above identify a wake-mediated collective instability that emerges only when elastic wakes generated by neighboring cantilevers interact. As discussed in Sec.~\ref{sec:flowfield}, this behaviour does not appear to be fully explained by the well-known elastic instability criterion of Pakdel and McKinley \citep{Mckinley1996,Pakdel1996}. Therefore, we now attempt to outline an alternative possible mechanism, accounting for the FSI, and based on the current experimental results along with insights taken from previous related works \citep{Varchanis2020,Haward2020, Hopkins2020,Hopkins2021, Haward2021}.

A possible mechanism is as follows. In general (for both the WS and HS fluids), at low Weissenberg numbers ($\text{Wi} \ll \text{Wi}^{*}$), elastic wakes of stretching polymer develop symmetrically behind each cantilever. As $\text{Wi}$ increases, in the absence of significant shear-thinning (i.e., for the WS fluid), the growth of elastic stresses within these wakes renders them increasingly susceptible to lateral perturbations and fluctuations \citep{Haward2019, Varchanis2020}. For a single cantilever, such perturbations remain statistically symmetric after time averaging, consistent with previous observations for isolated fixed cylinders  \citep{Varchanis2020,Haward2020,Haward2021}. In arrays, however, adjacent wakes can interact. Once neighboring wakes come into sufficiently close proximity, they merge into a common low-velocity region downstream of the array. The resulting blockage redistributes the incoming flow around the outer sides of the cantilevers, thereby establishing the divergent-flow state quantified by the flow divergence parameter. The associated change in the local stress distribution then provides a plausible route for the inward spanwise deflection of the outer cantilevers, observed in Fig.~\ref{fgr:pillar-relative_Newtonian}(e,f).

When shear-thinning effects are strong (i.e., for the HS fluid), they might be expected \textit{a priori} to promote divergent or asymmetric flow states around the cantilever arrays, consistent with previous studies \citep{Dey2018,Haward2019,Hopkins2020,Haward2020,Varchanis2020,Hopkins2021,Khan2021,Haward2021,Spyridakis2024}. However, several geometric features of the present system may contribute to the suppression of the merged-wake state for the HS fluid. First, the low blockage ratio ($B_R=0.008$) weakens in-plane flow constriction around each cantilever in the $x$--$y$ plane, reducing local shear-rate difference on either side of the cantilever. Second, the shallow channel with low aspect ratio ($A_R=H/W=0.1$) imposes a Poiseuille-like velocity profile in the $z$-direction with the strong velocity gradient between the channel floor and ceiling. The narrow gap between the cantilever tips and the channel ceiling further promotes this shear layer in the same direction. Since the shear-thinning fluid exhibits a reduced viscosity in high shear rate region, this velocity gradient is amplified relative to a Newtonian fluid \citep{lopez_2025}. This may enhance out-of-plane shear while reducing the tendency for in-plane wake interaction and gap closure. This provides a plausible explanation for why the HS fluid produces predominantly streamwise cantilever deflection while suppressing the spanwise deflection and wake merger observed for the WS fluid. 

As a result, the flow divergence and streamline curvature are not sufficiently developed in the $x$--$y$ plane to trigger the elastic instability, explaining the absence of the asymmetric flow states reported in previous studies of deformable cylinders in shear-thinning viscoelastic flows \citep{Dey2018,Hopkins2020}. If the outlined mechanism is correct, then it comes with some surprises given that $N_1$ is significantly greater for the more shear-thinning HS fluid than for the WS fluid. Additionally, it means that the shear-thinning effects described may counteract the effects of $N_1$ by maintaining a constant intercantilever gap as $\mathrm{Wi}$ is increased.

The flexibility of the cantilevers may further modify the instability threshold. Recently, \citet{Chandrashekar2025} reported the effect of geometric asymmetry on two-dimensional viscoelastic flow past a single cylinder. They showed that breaking the geometric symmetry by moving the cylinder away from the channel centerline results in an enhancement of the flow rate difference on either side of the cylinder for a weakly shear-thinning fluid, while reducing it for a highly shear-thinning fluid. These findings are qualitatively consistent with the behaviour observed in the present study. Therefore, if the flow is sensitive to spanwise deflection, transient asymmetry introduced by cantilever deformation may promote the same trends reported by \citet{Chandrashekar2025}, potentially contributing to the suppression of elastic instability in the HS fluid. On the other hand, this idea runs counter to observations made on single flexible cantilevers in shear-thinning viscoelastic flows, which showed rather strong spanwise displacements at elevated $\mathrm{Wi}$ \citep{Dey2018,Hopkins2020}.  %Even small spanwise deflections introduce geometric asymmetry into the flow.
The present results therefore indicate that the onset of the wake-mediated collective instability is governed not only by fluid rheology, but also by the interplay between elastic wake interactions, geometric confinement, and structural compliance.

\section{Conclusion}

In this study, we investigated viscoelastic fluid--structure interaction in side-by-side cantilever arrays using a bottom-up approach that systematically varied both the array geometry and the rheological properties of the test fluid. By comparing weakly shear-thinning (WS, $c=0.1$ wt\%) and highly shear-thinning (HS, $c=1.0$ wt\%) PEO solutions, we identified a wake-mediated collective instability that emerges only through interactions between neighboring elastic wakes. The simultaneous onset of wake merger, divergent flow, and inward spanwise cantilever deflection demonstrates that these flow and structural transitions are coupled manifestations of the same collective fluid--structure interaction process.

For the WS fluid, neighboring elastic wakes merge beyond a critical Weissenberg number, producing a common low-velocity region downstream of the array, redistributing the incoming flow, and driving coordinated inward spanwise deflection of the cantilevers. The critical Weissenberg number increases with the number of cantilevers, indicating that the onset of the collective instability is governed by the array geometry. In contrast, the HS fluid suppresses wake merger and consequently the associated divergent-flow state and coordinated spanwise structural response, despite comparable levels of fluid elasticity at equivalent $\mathrm{Wi}$. These observations demonstrate that fluid elasticity alone is insufficient to trigger the instability; rather, the collective transition depends on the ability of neighboring elastic wakes to interact and merge, a process strongly influenced by shear-thinning.

Our findings extend previous studies demonstrating wake-mediated interactions between upstream and downstream cantilevers by showing that elastic wakes can also couple neighboring cantilevers arranged side-by-side. More broadly, they establish elastic wake interactions as the physical mechanism linking local viscoelastic flow structures to the collective response of flexible cantilever arrays. These results advance the fundamental understanding of viscoelastic fluid--structure interactions in biological and bioinspired systems, including ciliary and filamentous arrays, and provide a framework for future investigations into how fluid rheology, geometric confinement, and structural compliance together govern collective transport and deformation in viscoelastic flows.

\vspace{+0.5cm}

\backsection[Acknowledgement]{We are grateful to Dr. Charlotte De Blois (Universit\'{e} Paris-Saclay) for her invaluable advice on fabrication techniques and image analysis.}

\backsection[Funding]{We gratefully acknowledge the support of the Okinawa Institute of Science and Technology Graduate University (OIST) with subsidy funding from the Cabinet Office, Government of Japan, and we also gratefully acknowledge Kakenhi funding from the Japan Society for the Promotion of Science (Grant Nos. 24K07332, 24K00810, and 25KJ2241).}

\backsection[Data availability statement]{The data that support the findings of this study are available within the article and appendix.}

\backsection[Declaration of interests]{The authors report no conflict of interest.}

\backsection[Author contributions]{Conceptualization, A.Y., A.Q.S., and S.J.H.; methodology, A.Y.; validation, A.Y.; formal analysis, A.Y. and S.J.H.; investigation, A.Y.; resources, A.Q.S.; data curation, A.Y.; writing—original draft preparation, A.Y.; writing—review and editing, A.Y., A.Q.S., and S.J.H.; supervision, A.Q.S., and S.J.H.; project administration, A.Q.S.; funding acquisition, A.Y., A.Q.S., and S.J.H.}

\vspace{+6cm}

\begin{appendix}

%\appendix
\setcounter{figure}{0}
\renewcommand{\figurename}{Fig.}
\renewcommand{\thefigure}{A\arabic{figure}}
%\section{}\label{appA}
%\appendix
\begin{figure*}
\centering
  \includegraphics[scale=0.35]{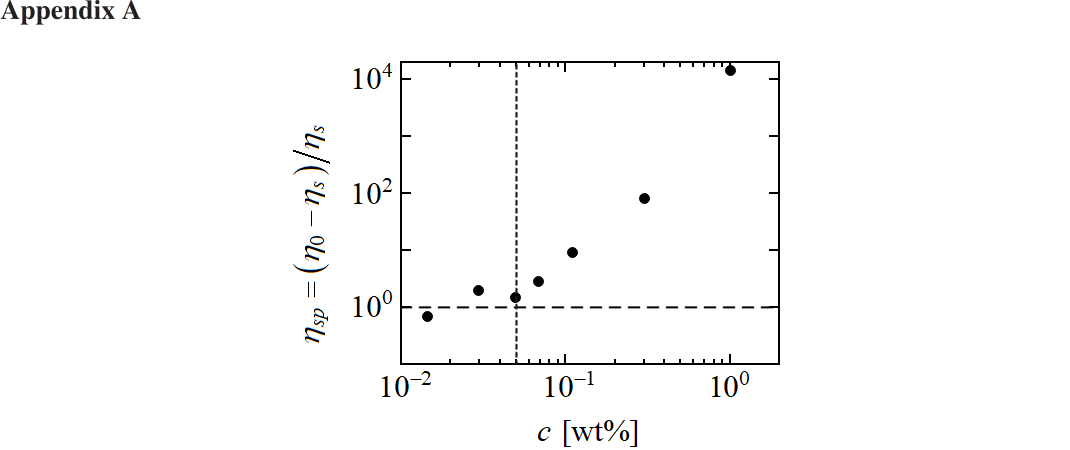}
  \caption{Specific viscosity ($\eta_{sp}=(\eta_{0}-\eta_{s})/\eta_{s}$ as a  function of polymer concentration $c$. The overlap concentration $c^* \approx 0.05$~wt\% is decided from the point where the specific viscosity begins to increase above $\eta_{sp}=1$ (as marked by the vertical dotted line).
}
  \label{fgr:Hspvsc}
\end{figure*}

\begin{figure*}
\centering
  \includegraphics[scale=0.42]{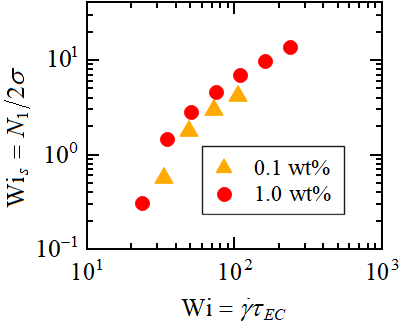}
  \caption{Weissenberg number determined from normal and shear stress measurements made in steady shear ($\text{Wi}_s = N_1/2\sigma$) plotted against the Weissenberg number defined as a product of the steady shear rate and the elastocapillary thinning time ($\text{Wi}=\dot\gamma \tau_{EC}$), showing a data collapse and a consequent one-to-one correspondence between $\mathrm{Wi}$ and $\text{Wi}_s$ for both polymeric test solutions at $c = 0.1$ and 1.0 wt\%.}
  \label{fgr:N1vsWi}
\end{figure*}

\begin{figure*}
\centering
  \includegraphics[scale=0.42]{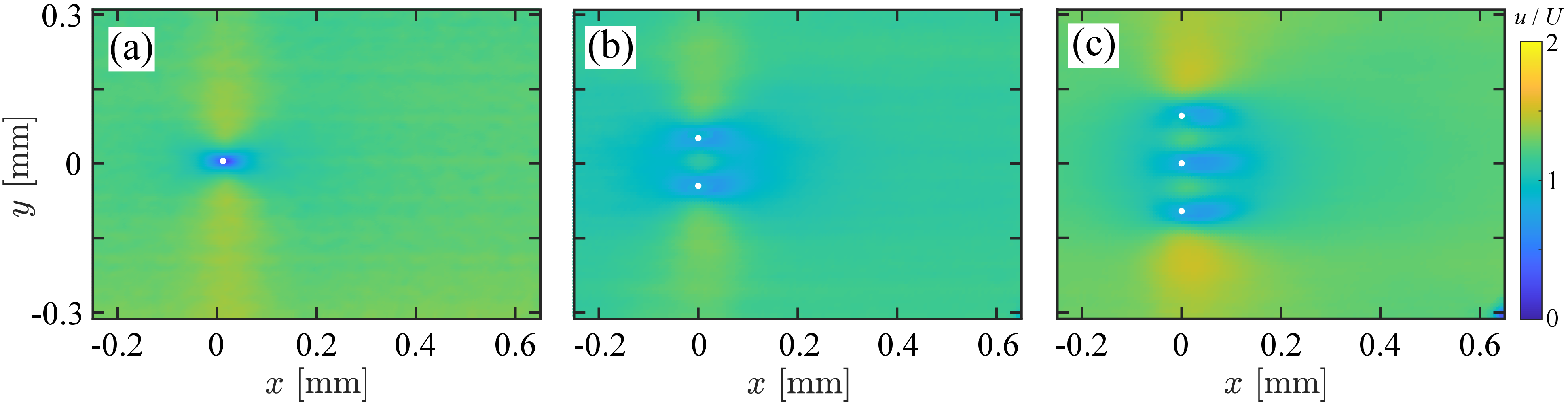}
  \caption{Representative time-averaged normalized flow fields for flow of the Newtonian reference fluid (45~wt\% sodium iodide aqueous solution) around the various cantilever configurations at $\text{Re} \approx 0.25$: (a) single cantilever, (b) double cantilever, (c) triple cantilever. Flow is from left to right.}
  \label{fgr:newtonian_table}
\end{figure*}

\begin{figure*}
\centering
  \includegraphics[scale=0.5]{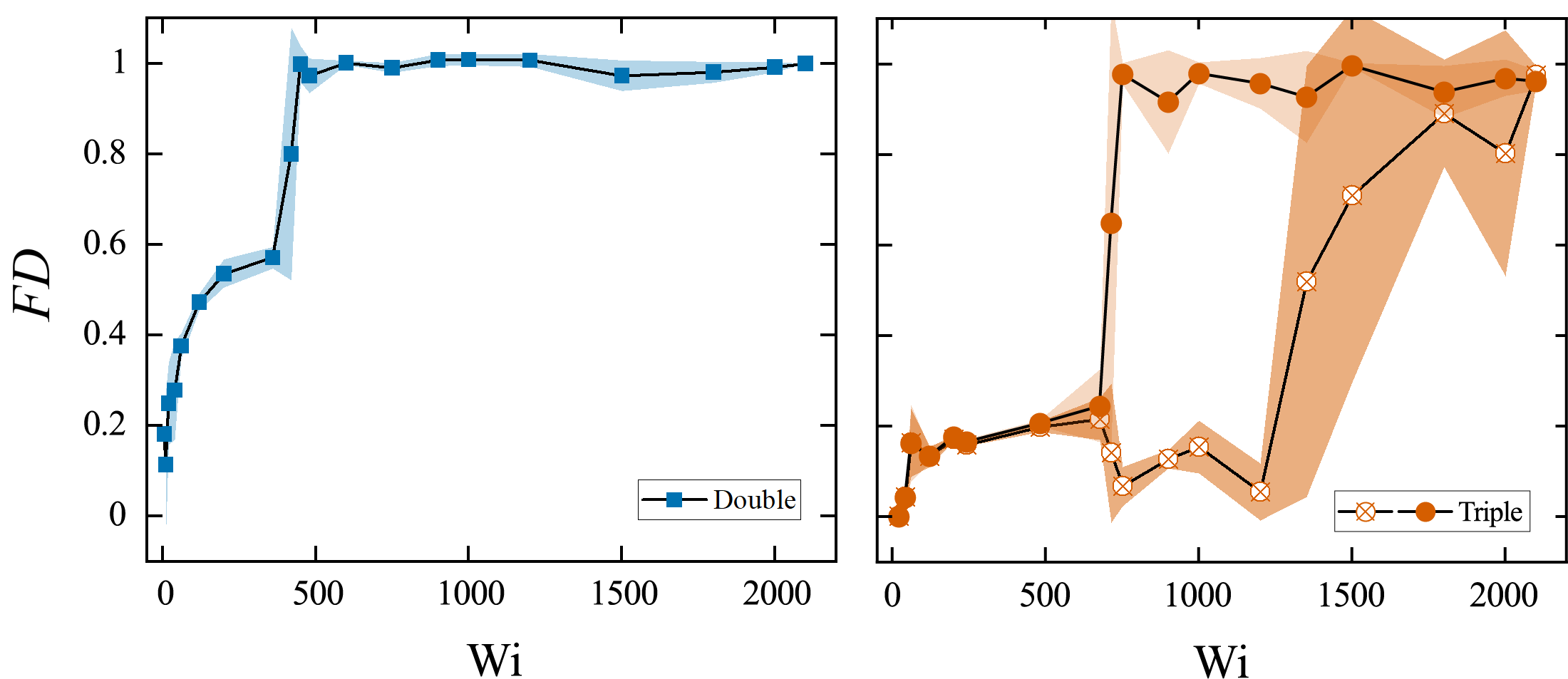}
  \caption{Flow divergence ratio $FD$ (Eq.~\ref{eq:divergent}) as a function of the Weissenberg number Wi for the weakly shear-thinning fluid ($c = 0.10$~wt\%) and the double and triple cantilever configurations. Error bounds represent the full range of values measured over twice-repeated experiments.
}
  \label{fgr:I_error}
\end{figure*}

\begin{figure*}
\centering
  \includegraphics[scale=0.35]{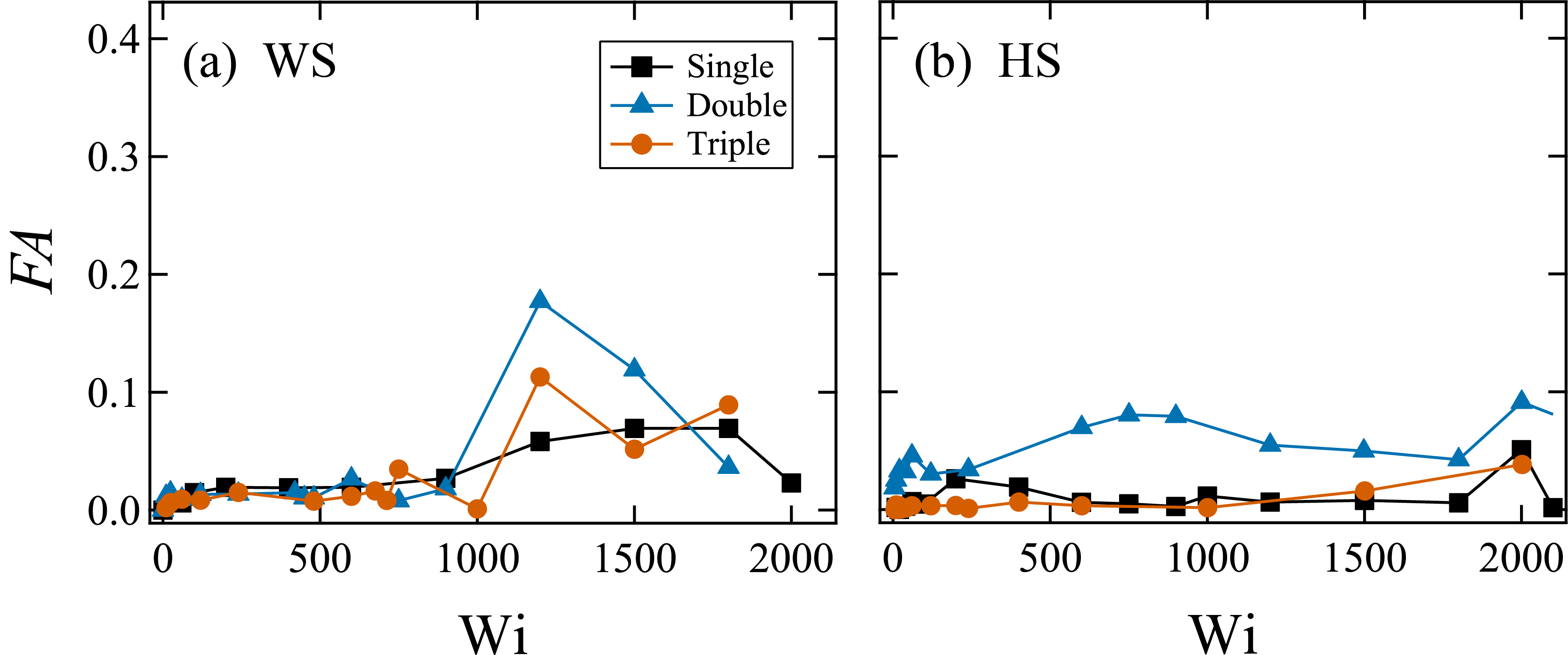}
  \caption{Flow asymmetry ratio $FA$ (Eq.~\ref{eq:asymmetry}) as a function of the Weissenberg number $\mathrm{Wi}$ with different number of cantilevers (single, double and triple) for (a) the weakly shear-thinning (WS) fluid, and (b) the strongly shear-thinning (HS) fluid. 
}
  \label{fgr:I_star}
\end{figure*}

\begin{figure*}
\centering
  \includegraphics[scale=0.22]{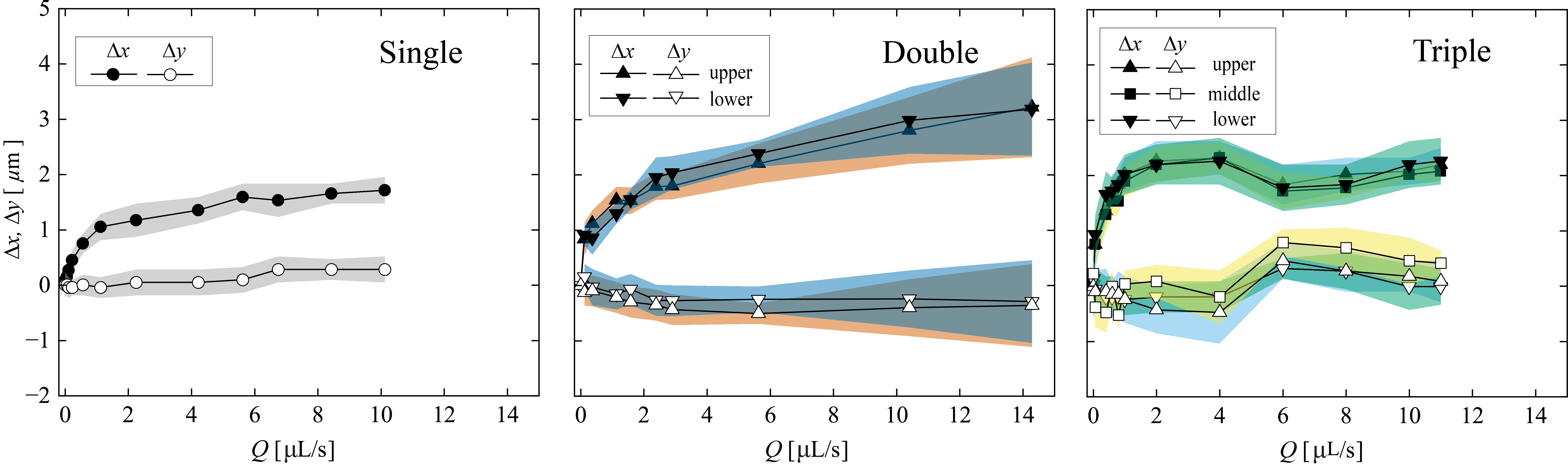}
  \caption{Cantilever displacement in the streamwise ($\Delta x$) and spanwise ($\Delta y$) directions for Newtonian fluid (70~wt\% glycerol/water mixture) with different cantilever configurations (single, double, triple). The shaded regions represent the time variation of the cantilever deflection, which is found from the width of the probability density distribution at 10~\% of its peak value.}
    \label{fgr:Newtonian_pillardis}
\end{figure*}

\end{appendix}
\clearpage

%\bibliographystyle{jfm}
%\bibliography{jfm}

\providecommand{\noopsort}[1]{}\providecommand{\singleletter}[1]{#1}%

%Use of the above commands will create a bibliography using the .bib file. Shown below is a bibliography built from individual items.

%% End of file `jfm.bib'.

\end{document}